\documentclass[10pt]{article}

\usepackage{amsmath}
\usepackage{amssymb}

\usepackage{graphicx}

\usepackage{cite}

\usepackage{color} 

\usepackage[labelfont=bf,labelsep=period,justification=raggedright]{caption}

\makeatletter
\renewcommand{\@biblabel}[1]{\quad#1.}
\makeatother

\date{}

\newcommand{\e}{\mathrm{e}}             

\newcommand{\eg}[1]
  {{\it e.g.\/}\ifx#1.\else\expandafter#1\fi}
\newcommand{\ie}[1]
  {{\it i.e.\/}\ifx#1.\else\expandafter#1\fi}
\newcommand{\insilico}{{\it in silico\/}}

\newcommand{\invitro}{{\it in vitro\/}}

\newcommand{\cell}{{\rm cell}}       
\newcommand{\cm}{\ensuremath{\mathrm{cm}}}
\newcommand{\Min}{\ensuremath{\mathrm{min}}}
\newcommand{\mm}{\ensuremath{\mathrm{mm}}}

\newcommand{\um}{\ensuremath{\mu\mathrm{m}}}
\newcommand{\mV}{\ensuremath{\mathrm{mV}}}
\newcommand{\s}{\ensuremath{\mathrm{s}}}

\renewcommand{\@}{\partial}             
\newcommand{\Complex}{\mathbb{C}}       
\newcommand{\crit}{\mathrm{crit}}       
\renewcommand{\d}{{\mathrm d}}             
\newcommand{\Df}[2]{\frac{\d{#1}}{\d{#2}}} 
\newcommand{\Heav}{H}                   
\renewcommand{\Im}[1]{{\rm Im}\ifx#1$ $\else\expandafter\left(#1\right)\fi}	
\renewcommand{\i}{\mathrm{i}}           
\newcommand{\inner}[2]
  {\left\langle#1\, , \,#2\right\rangle}
\newcommand{\mx}[1]{\mathbf{#1}}        
\renewcommand{\Re}[1]{{\rm Re}\ifx#1$ $\else\expandafter\left(#1\right)\fi}	
\newcommand{\Real}{\mathbb{R}}          
\newcommand{\T}{^{\mathrm{T}}}          

\newcommand{\Ampl}{A}                   
\newcommand{\angd}{\theta}              
\renewcommand{\c}{c}                    
\newcommand{\curv}{\kappa}              
\newcommand{\D}{\mx{D}}                 
\newcommand{\Dfac}{D}                   
\newcommand{\dfdp}{\@_{\param}\f}       
\newcommand{\distd}{\rho}               
\newcommand{\distdeq}{\distd_{*}}       
\newcommand\dtime{\dot }                
\newcommand{\F}{F}                      
\newcommand{\Fa}{s}                     
\newcommand{\Fc}{a}                     
\newcommand{\f}{\mx{f}}                 
\newcommand{\Grad}{G}                   
\newcommand{\Gradm}{g}                  
\newcommand{\Gradang}{\phi}             
\newcommand{\h}{\mx{h}}                 
\newcommand{\n}{\mx{n}}                 
\renewcommand{\O}{\mathcal{O}}          
\newcommand{\param}{\alpha}                  
\newcommand{\paravg}{\overline{\param}} 
\newcommand{\parmax}{\param_{\max}}
\newcommand{\R}{{\vec R}}               
\newcommand{\Rc}{\vec{r}_c}             
\newcommand{\rc}{r_c}                   
\newcommand{\Ri}{R_i}                   
\newcommand{\RF}{\mx{W}}                
\renewcommand{\r}{{\vec r}}             
\newcommand{\sens}{\gamma}              
\newcommand{\senscurv}{\sens_{\curv}}   
\newcommand{\sensr}{\sens_r}            
\newcommand{\hr}{\h_r}                  
\newcommand{\epsr}{\epsilon_r}          
\newcommand{\sensp}{\sens_{\param}}     
\newcommand{\epsp}{\epsilon_{\param}}   
\newcommand{\sensi}{\sens_i}            
\newcommand{\epsi}{\epsilon_i}          
\newcommand{\sensD}{\sens_D}            
\newcommand{\hD}{\h_D}                  
\newcommand{\epsD}{\epsilon_D}          
\newcommand{\U}{\mx{U}}                 
\renewcommand{\u}{\mx{u}}               
\newcommand{\Xc}{x_c}                   
\newcommand{\Zc}{z_c}                   

\newcommand{\V}{V}

\newcommand{\m}{m}

\newcommand{\fgate}{f}

\newcommand{\ik}{i_{K_1}}
\newcommand{\ix}{i_{x_1}}
\newcommand{\ina}{i_{Na}}
\newcommand{\is}{i_s}
\newcommand{\Iext}{I_{ext}}

\newcommand{\alpm}{\alpha_m}
\newcommand{\betm}{\beta_m}

\newcommand{\Cm}{C_m}

\newcommand{\gna}{g_{Na}}

\newcommand{\gs}{g_s}

\newcommand{\alpavg}{\overline{\alpha}}
\newcommand{\alposc}{\alpha_{\mathrm{osc}}}

\newcommand{\bwidth}{w}

\newcommand{\delpar}{\delta_{\param}}
\newcommand{\Dmin}{D_{\min}}
\newcommand{\Dmax}{D_{\max}}
\newcommand{\noise}{\eta}

\newcommand{\Nx}{N_x}                   
\newcommand{\Ny}{N_y}                   
\newcommand{\Nz}{N_z}                   

\newcommand{\eqlabel}[1]{\label{eq:#1}} 
\newcommand{\eq}[1]{(\ref{eq:#1})}      
\def\eqtwo(#1,#2){(\ref{eq:#1},\ref{eq:#2})}  
\def\eqset(#1,#2){(\ref{eq:#1}--\ref{eq:#2})}  
\newcommand{\Fig}[1]{Figure~\ref{#1}}     
\newcommand{\fig}[1]{figure~\ref{#1}}     
\newcommand{\figs}[1]{figures~\ref{#1}}   
\newcommand{\figref}[1]{\ref{#1}}       

\newcommand{\myfigure}[3]
  {\begin{figure*}[tbp]\centerline{\includegraphics{#1}}\caption[]{#2}\label{#3}\end{figure*}}
\newcommand{\titfigure}[4]
  {\begin{figure*}[tbp]
      \centerline{\includegraphics{#1}}
      \caption[#2]{\textbf{#2} #3}
      \label{#4}
    \end{figure*}}

\begin{document}


\begin{flushleft}
{\Large
\textbf{Evolution of spiral and scroll waves of excitation \\
 in a mathematical model of ischaemic border zone}
}
\\
Vadim N. Biktashev$^{1,\ast}$, 
Irina V. Biktasheva$^2$,
Narine A. Sarvazyan$^3$
\\
\bf{1} Department of Mathematical Sciences, University of Liverpool, 
   Liverpool, UK
\\
\bf{2} Department of Computer Science, University of Liverpool, 
   Liverpool, UK
\\
\bf{3} Pharmacology and Physiology Department, The George
  Washington University, Washington DC, USA
\\
$\ast$ E-mail: Corresponding vnb@liv.ac.uk
\end{flushleft}

\section*{Abstract}

Abnormal electrical activity from the boundaries of ischemic cardiac
tissue is recognized as one of the major causes in generation of
ischemia-reperfusion arrhythmias. Here we present theoretical analysis
of the waves of electrical activity that can rise on the boundary of
cardiac cell network upon its recovery from ischaemia-like
conditions. The main factors included in our analysis are macroscopic
gradients of the cell-to-cell coupling and cell excitability and
microscopic heterogeneity of individual cells. The interplay between
these factors allows one to explain how spirals form, drift together
with the moving boundary, get transiently pinned to local
inhomogeneities, and finally penetrate into the bulk of the
well-coupled tissue where they reach macroscopic scale.  The
asymptotic theory of the drift of spiral and scroll waves based on
response functions provides explanation of the drifts involved in this
mechanism, with the exception of effects due to the discreteness of
cardiac tissue. In particular, this asymptotic theory allows an
extrapolation of 2D events into 3D, which has shown that cells within
the border zone can give rise to 3D analogues of spirals, the scroll
waves. When and if such scroll waves escape into a better coupled
tissue, they are likely to collapse due to the positive filament
tension. However, our simulations have shown that such collapse of
newly generated scrolls is not inevitable and that under certain
conditions filament tension becomes negative, leading to scroll
filaments to expand and multiply leading to a fibrillation-like state
within small areas of cardiac tissue.

\section*{Introduction}

Heart is a remarkably reliable machine whose function is to pump the
blood as required by the organism. An important part of its work is
the orderly propagation of electrical signal, that is the wave of
excitation passing through cardiac muscle, which subsequently triggers
its ordered contraction. Abnormalities of the excitation wave
propagation, known as arrhythmias, are precursors of sudden cardiac
arrest and other life-threatening pathologies.   This
  paper focuses on mathematical analysis of arrhythmogenic conditions
  associated with cardiac tissue recovery from acute ischemia, also
  known as \emph{reperfusion arrhythmias}. Such recovery can be more
  dangerous then ischemia itself and often leads to ventricular
  fibrillation and sudden cardiac death~\cite{Wit-Janse-2001}.
  Reperfusion can be spontaneous (relief of coronary spasm, dislodging
  of a thrombus) or externally imposed (antithrombolitic therapy,
  angioplasty). It can also occur on a microscopic scale during
  ischemia itself as a result in shifts in
  microcirculation~\cite{Kay-etal-2008}.
  As of today, the exact mechanisms of reperfusion arrhythmias
  remain poorly understood. 
This is because the inner
layers of ischaemic boundary are inaccessible for live visualization
on a spatial scale required to distinguish behaviour of individual
cells. Therefore, in order to understand how the abnormal activity
spreads from single cells to the bulk of cardiac tissue, we and others
had to rely on either \invitro\ experimental preparations or on computer
modeling.

Our work builds on the experimental data acquired from monolayers of
cardiac myocytes under conditions that mimicked the ischaemic
boundary~\cite{%
  Arutunyan-etal-2001,%
  Arutunyan-etal-2002,%
  Arutunyan-etal-2003%
}, and the results of direct numerical simulations that closely
matched these experimental observations~\cite{%
  Arutunyan-etal-2003,%
  Pumir-etal-2005,%
  Biktashev-etal-2008%
}.  The \emph{in silico} modelling provided an explanation to several
experimental findings, including the dependence of drift of
boundary-bound spirals on their chirality, pin-drift-pin type of
spiral tip motion and the effect of boundary movement on spiral
detachment~\cite{%
  Pumir-etal-2005,%
  Biktashev-etal-2008%
}.

\titfigure{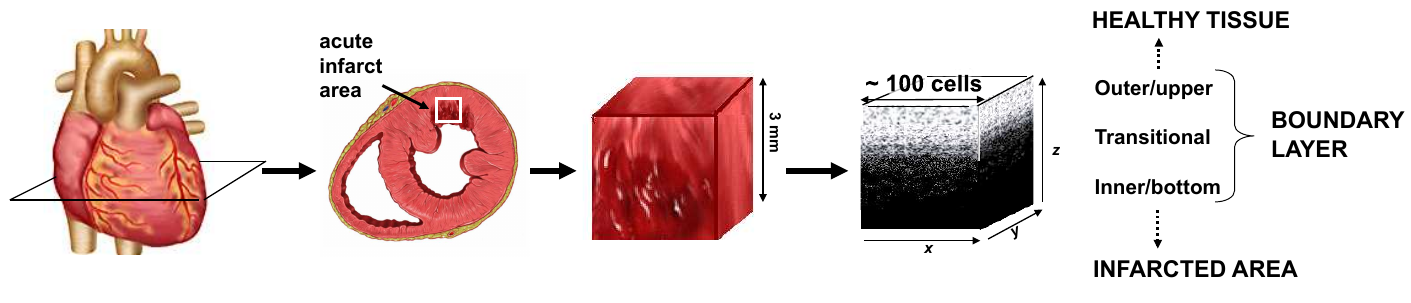}{
  We consider excitation dynamics 
  on a microscopic spatial scale, in areas of cardiac tissue with
    severely suppressed cell-to-cell coupling 
    superimposed with elevated cell excitability. 
}{ 
}{microscopic}

The rotating waves of activity to be discussed in this paper, occur on
a much smaller spatial scale as compared to classical cardiac
reentry~\cite{%
  Pertsov-etal-1993,%
  Gray-etal-1995,%
  Chen-etal-1997,%
  Efimov-1997,%
  Weiss-etal-2005%
}, see~\fig{microscopic}. Specifically, we are focusing on a
dynamically and spatially changing set of conditions which can occur
within a thin layer of cells sandwiched between intact healthy tissue
and the recovering ischaemic areas. Myocytes within such layers can
become spontaneously active as a result of calcium overload and/or
local noradrenaline release.  The impact of intrinsic myocyte
heterogeneity on network behaviour is markedly enhanced due to
decrease in electrical coupling between the cells. It gets even more
complicated as the physicochemical factors that create the boundary,
such as low pH, lack of oxygen, hyperkalemia, noradrenaline, move in
space due to the dynamic nature of reperfusion.  Altogether the moving
boundary, heterogeneous substrate, steep gradient of coupling and
self-oscillatory activity of individual cells can give rise to a rich
network behaviour discussed in our previous
paper~\cite{Biktashev-etal-2008}. A continuous generation of
mini-reentries from individual ectopic sources occurs within the least
coupled cells layers, and then the activity spreads towards the better
coupled layers of the boundary~[8].  This scenario was suggested by
our experiments in neonatal rat cardiomyocytes and was later supported
and expanded upon using the \insilico\ approach.  Yet, numerical
modelling of cellular behaviour has its limitations, and there is a
need to understand how much of the phenomena observed in the
simulations are generic and how much of it depends on the specifics of
the model. Further still, cardiac tissue is three-dimensional, whereas
our experiments and simulations reported previously were conducted
using two-dimensional cell networks. Extrapolation of the two
dimensional data into three dimensions requires additional theoretical
understanding.

In the present paper, we use an asymptotic theory of spiral and scroll
waves' drift together with the recently developed numerical technique
to compute the response functions of spiral waves \cite{%
  Biktasheva-Biktashev-2003,%
  Biktasheva-etal-2009,%
  Biktasheva-etal-2010%
} to provide theoretical analysis of our experimental and numerical
data.  We then use this theoretical framework to predict behaviour of
the scroll waves in an ischaemic border zone in 3D,
where such experiments are not currently feasible.
Finally we confirm theoretical 3D predictions by numerical simulations
of cell network behaviour.

Specifically, we address the following questions:
\begin{enumerate}
\item In both experiments and numerical simulations, spiral waves
were not static within the border zone. What determines the
components of the drift velocity, and why the spiral cores
can be dragged together with the moving border zone?%

\item In both experiments and numerical simulations, the drift of the
spirals was interrupted by their ``pinning'' to clusters of cells. We
have shown numerically that these can be cell clusters of either
elevated or suppressed excitability. What is the mechanism of such
pinning?%

\item In both experiments and numerical simulations, the episodes of
spiral drift and pinning alternated. What is the mechanism by which
pinning can give way to further drift?%

\item One of arrhythmogenic scenarios proposed
  in~\cite{Pumir-etal-2005,Biktashev-etal-2008} involved pinning of a
spiral wave to a local heterogeneity which persists long enough until
the border zone passes and the spiral gets into the better coupled
tissue. Is this scenario viable in 3D?%

\end{enumerate}

\section*{Methods}

\subsection*{Direct Numerical simulations: tissue model}

\titfigure{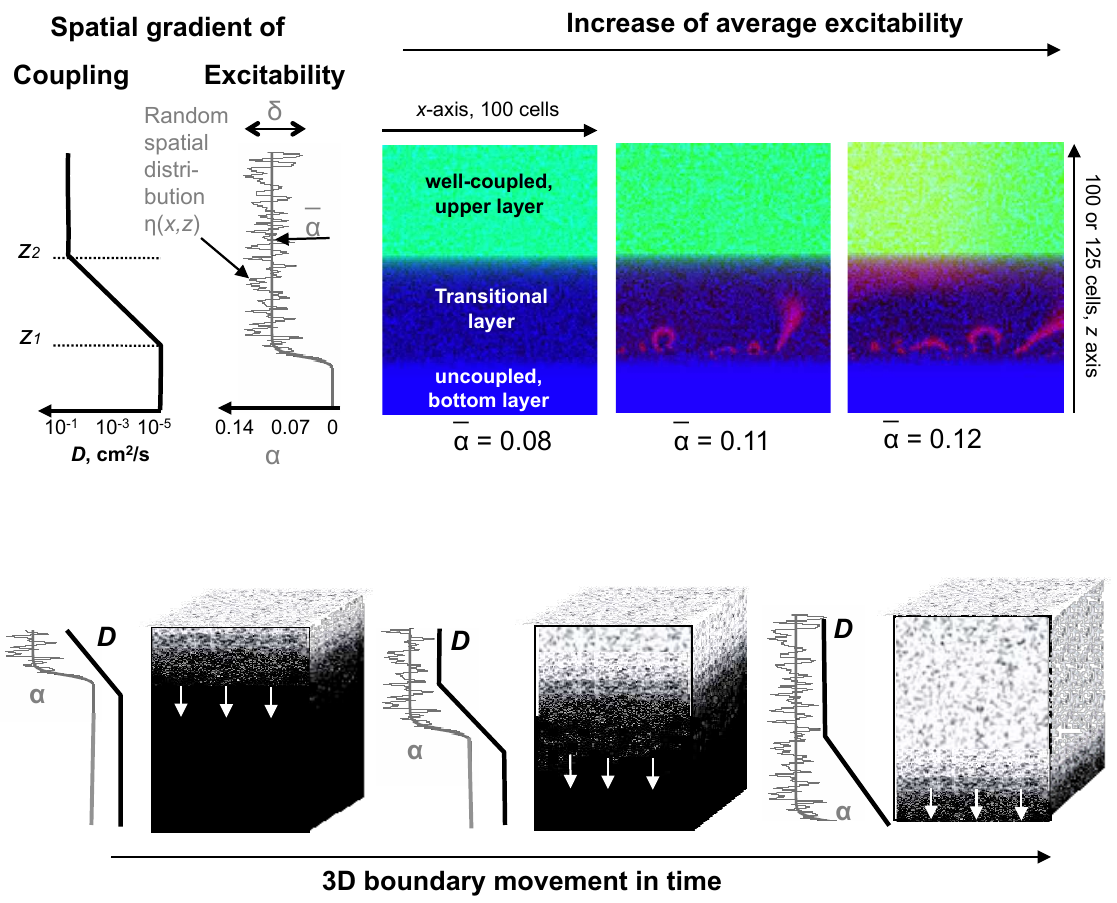}{
  Schematic of numerical protocols.
}{  
  Top row: 2D setting~\cite{Biktashev-etal-2008}. Distribution of the
  diffusivity $\Dfac$ and excitability/automaticity $\param$ across
  the border zone. The three colour panels are representative
  snapshots of solutions at different values of $\alpavg$, as it was
  slowly growing at a fixed profile of $\Dfac$.  Here and below we use
  the red colour component to show the excitation wave (transmembrane
  voltage), blue component for the cell excitability/automaticity
  (denoted as $\param$, see definition in the text) and the green
  component for the cell electrical coupling strength (denoted as
  $\Dfac$ for transmembrane voltage diffusivity).  E.g. yellow is a
  sum of green and red, and magenta is a sum of red and blue.
  Bottom row: 3D setting for this paper. The transition zone moves downwards.
}{schematic}

\titfigure{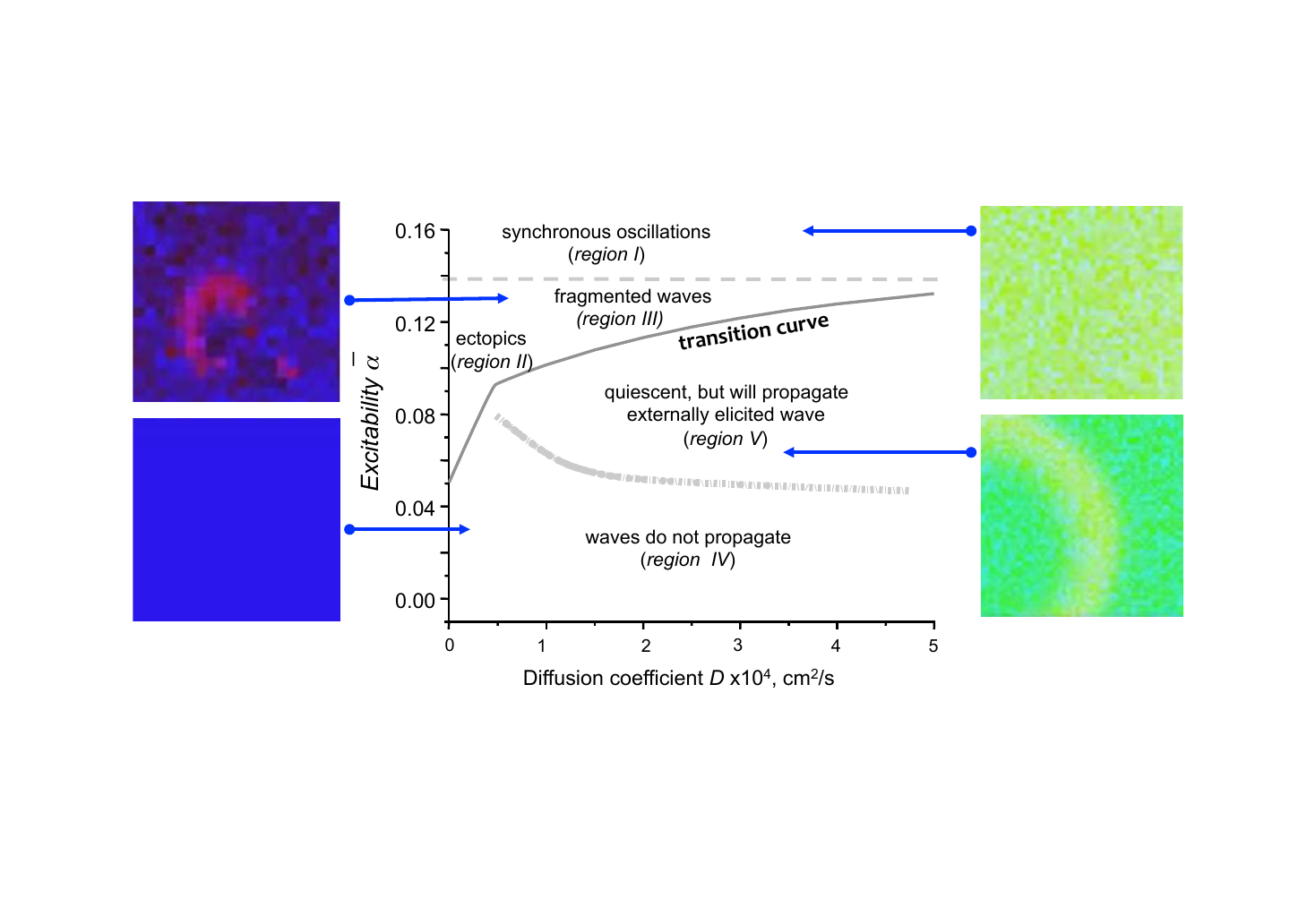}{
  {The parameter space diagram of the numerical
  model~\eqtwo(brp,heterogeneity)~\cite{Pumir-etal-2005}.}
}{
  {The parameter regions I-V correspond to distinctive
  regimes of wave initiation and propagation, observed in simulations
  where $\paravg$ and $\Dfac$ were maintaned constant throughout the
  simulations.  The panels on the sides show representative snapshots
  of solutions, corresponding to regions I, III, IV and V. }
}{diagram}


The mathematical model mimicking the conditions 
  when tissue recoveres from acute
  ischaemia, and its experimental foundations are described in detail
  in our previous works~\cite{Biktashev-etal-2008} and references
  therein. To capture the complexity of
  pathophysiological conditions associated with reperfusion
  arrhythmias, we use a simplified kinetic model of individual
  cells, and enrich it
  by adding individual cell heterogeneity, different course of
  recovery of cell coupling and excitability, and spatial
  arrangement of conditions on the boundary of ischemic tissue.
The importance of the latter three factors, cell heterogeneity,
  individual cell excitability and cell-to-cell coupling, for the
  cardiac network behaviour was studied in our previous
  paper~\cite{Pumir-etal-2005}. The arguments and experimental
  evidence presented there suggest that from the network/tissue perspective,
  it is not very important exactly how these properties are
  altered. In this paper, we model a tissue recovering from acute ischaemia as
  a three-layered slab made of a heterogeneous mix of
  cells, subject to a vertical gradient of average cell excitability and
  a vertical gradient of cell-to-cell coupling strength. 
  The 2D and 3D versions of the model are illustrated
  in~\fig{schematic}.  %
In terms of the
parametric diagram described in~\cite{Pumir-etal-2005} and shown
in~\fig{diagram}, the bottom layer corresponds to the parametric
region IV. It has low excitability and weak coupling which result in
the quiescent state where propagation is not possible.  The outer
layer with high excitability and strong coupling is in the parametric region V of
the digram, corresponding to the quiescent state where wave
propagation is possible.  The middle, or transitional, layer is
sandwiched between inner and outer layers, so, from bottom to top, it
starts in region III (high excitability and weak coupling resulting in
spontaneous fragmented waves) and then via a gradual increase in
coupling strength proceeds to region V (high excitability, strong
coupling, quiescent state where wave propagation is possible)
characteristic of the upper layer. The layers are not static but move
downwards through the slab, which represents the reperfusion, or
wash-out, of the agents affecting the relevant tissue properties.
Depending on type of reperfusion, blood flow can
  recover within seconds (cases of resolved coronary spasm,
  spontaneous dislodging of thrombi, angioplasty) or within minutes
  (cases of changes in coronary flow due to gradual accumulation of
  metabolites or pharmacological interventions). Therefore, the
  dynamics of moving border zone can vary in a rather wide range, from
  $\cm/\s$ to $\mm/\Min$. We select the values of the border zone
  speed that produce interesting effects. 

We assume that the cells are arranged in a rectangular grid of
$\Nx\times\Nz$ (in 2D) or $\Nx\times\Ny\times\Nz$ (in 3D) cells
connected to each other via Ohmic resistances. Properties of the cells
and resistivities of the contacts are varied in time an space.  The
cells are assumed to have linear size of $30\,\um$ which serves as a
space scale to endow the voltage diffusivity and other space-related
quantities with suitable dimensionality.  The cells are connected to
the nearest neighbours, so an internal cell has four contacts in 2D
and six contacts in 3D.

The excitable dynamics of cells is described by the
Beeler-Reuter-Pumir~\cite{Pumir-etal-2005} (BRP) model of a neonatal
cardiac myocyte. The BRP model is based on the generic
Beeler-Reuter~\cite{Beeler-Reuter-1977} model of a cardiac myocyte,
which contains an explicit, albeit simplified, description of
individual ionic currents, and was slightly modified to match the
ionic currents reported for neonatal cardiac cells used in our
experiments~\cite{Pumir-etal-2005}.  The complete set of the BRP model
equations is given the Appendix; here we only outline the
modifications. The affected equations are
\begin{eqnarray}
  \dot{\V}  &=& -(1/\Cm)(\ik+\ix+\ina+\is)+\Iext, \nonumber\\
  \m    &=& \alpm/(\alpm+\betm), \nonumber\\
  \ik   &=& 0.35 (0.3-\param) 
       \left[ \frac{4\,\e^{0.04 (\V+85)}-1}%
                     {\e^{0.08(\V+53)} + \e^{0.04
                         (\V+53)}} 
       + \frac{ 0.2 (\V+23)}{1-\e^{-0.04
            (\V+23)}} \right], \nonumber\\
      \Iext &=& \nabla\left( \Dfac(z,t) \nabla \V \right). \eqlabel{brp}
\end{eqnarray}
The last equation in~\eq{brp} is written, for brevity, as the
continuous limit, whereas actual calculations of the inter-cellular
currents were discrete, as described in more details in the
Appendix. The coupling strength between the cells is represented by
the voltage diffusion coefficient $\Dfac(z,t)$, and some of
the values of $\Dfac$ we use here are too low to hold the continuous
limit of the \eq{brp}. Note that as far as the continuous limit is
concerned, the voltage diffusivity $\Dfac$ is the only quantity in the
model related to space, so while within this limit, all results are
easily rescaled from one value of $\Dfac$ to another.

The maximum permeability of the fast inward current $\gna$ is 60\% of
the standard (2.4 vs 4), and that of the slow inward current, $\gs$,
is 50\% of the standard (0.045 vs 0.09).

We also have altered the balance between inward and outward currents
by inhibiting the inward potassium rectifier current, $\ik$
\cite{Pumir-etal-2005,Silva-Rudy-2003,Dhamoon-Jalife-2005}.
Suppression of $\ik$ to $30\%$ of the standard value mimics its
smaller contribution reported for neonatal
cardiomyocytes~\cite{Masuda-Sperelakis-1993,Wahler-1992} as compared
to the original Beeler and Reuter values for adult ventricular
cells~\cite{Beeler-Reuter-1977}. We use this supressed value of $\ik$
($\param=0$) for the bottom layer of the ischemic slab. In the upper
layers, further suppression, represented by the factor $(0.3-\param)$,
$\param>0$, enhances excitability. For high enough values of $\param$,
this leads to spontaneous firing of individual cells, \ie\ makes them
automatic~\cite{Pumir-etal-2005}.  
In~\cite{Pumir-etal-2005} we considered $\param$ values that led to
  the \insilico\ network
  behaviour closely matching the behaviour of neonatal cardiomyocyte
  layers. The excitability of the latter cells was increased using
  beta-adrenergic stimulation with
  isoproterenol~\cite{Arutunyan-etal-2003}
  and ischaemia-reperfusion protocol~\cite{Arutunyan-etal-2002}.
  Compared to~\cite{Pumir-etal-2005}, 
  here we only
  consider a narrow range of values of $\param$, where phenomena
  interesting for our present study are observed.
In our previous
paper~\cite{Biktashev-etal-2008}, parameter $\param=\param(x,y,z,t)$
varied in space and time and it was essential that it covered both
excitable and automatic regimes, so it was called both
``automaticity'' and ``excitability''. Here we concentrate mostly on
the events happening in the excitable regime ($\param\le0.13$, within
the range of intermediate coupling values, or region V in the
parametric space, shown in~\fig{diagram}), hence for brevity we mostly
refer to parameter $\param$ as ``excitability parameter'' or simply
``excitability''.  It should be kept in mind, however, that due to the
above ambiguity, this usage may differ from the meaning of
``excitability parameter'' in other studies.

Heterogeneity of individual cells' excitability is described as 
\begin{equation}
  \param(x,y,z,t)=\paravg(z,t)\,(1+\delpar\noise(x,y,z)),
                                                  \eqlabel{heterogeneity}
\end{equation}
where $\noise(x,y,z)$ is the Gaussian distributed uncorrelated random
variable with unit dispersion, and parameter $\delpar$ represents the
intensity of heterogeneity.

Space-time variations of $\Dfac$ and $\paravg$ are defined as
\begin{equation}
  \Dfac (z,t)=\left\{\begin{array}{ll}
    \Dmin, & z\leq z_1 , \\
    {\Dmin}^{\frac{z_2-z}{z_2-z_1}}\,{\Dmax}^{\frac{z-z_1}{z_2-z_1}}, 
	   & z_1\leq z\leq z_2 , \\
    \Dmax, & z \geq z_2, 
  \end{array}\right.                              \eqlabel{Dzt}
\end{equation}
and
\begin{equation}
  \paravg(z,t)=\frac12\left(1+\tanh\left(\frac{z-z_1}{\bwidth}\right)\right)\parmax,
                                                  \eqlabel{alpzt}
\end{equation}
where $z$ is the coordinate across the boundary, $z_1=z_1(t)$ and
$z_2=z_2(t)$ are the limits of the steepest part of the coupling
gradient, $\Dmax$ is the diffusion coefficient in the upper,
well-coupled layer, $\Dmin$ corresponds to the bottom, uncoupled
layer, and $\parmax$ is the highest level of excitability within the
slab.  We used the boundary width $\bwidth=3\times30\um$ in all
simulations.  Parameters $z_1$, $z_2$ vary linearly in time,
$z_1=z_{1,0}-\c t$, $z_2=z_{2,0}-\c t$.

Thus, the recovering ischaemic tissue is modelled as layers with
imposed excitability and coupling profiles as shown
in~\fig{schematic}. Specifically, we are modelling experimental
conditions when previously severely uncoupled ischaemic areas are
reperfused with agents which elevate cell excitability.

Finally, we also made simulations with deliberately arranged
parametric distributions not exploiting random number generators. The
details of those are given where the results are described.

\subsection*{Asymptotic theory of drift}
The asymptotic theory of spiral and scroll dynamics
under small perturbations~\cite{%
  Biktashev-1989,%
  Biktashev-Holden-1995,%
  Biktasheva-Biktashev-2003,%
  Biktasheva-etal-2010,%
  Foulkes-etal-2010%
} is formulated for the ``reaction-diffusion'' system of partial
differential equations (PDEs),
\begin{equation}
\@_t\u = \f(\u) + \D \nabla^2 \u  + \epsilon \h, \quad 
\u,\f,\h\in\Real^\ell,\; 
\D\in\Real^{\ell\times\ell},\;
\ell\ge2,					\eqlabel{RDS}
\end{equation}
where $\u(\r,t)=(u_1,\dots u_{\ell})\T$ is a column-vector of the
reagent concentrations, $\f(\u)=(f_1,\dots f_{\ell})\T$ is a
column-vector of the reaction rates, $\D$ is the matrix of diffusion
coefficients, $\r\in\Real^m$ ($m=2$ or $3$) is the vector of
coordinates, and $\epsilon\h=\epsilon \h(\u;\r,t)$ is some small
perturbation of the right-hand side, $|\epsilon|\ll1$.  For the
Beeler-Reuter-Pumir model, $\ell=7$, and $\D=\Dfac\n$, where
$\n=[n_{i,j}]$, $n_{1,1}=1$ and $n_{i,j}=0$ otherwise.

The theory assumes that spiral wave solutions to 
  equations~\eq{RDS} for $m=2$ are stationary rotating, not
  meandering. This is indeed satisfied for BRP model for all
  $\param$ values considered. Mathematically, the assumption
  means that 
  a spiral wave solution to \eq{RDS} for $m=2$ in the $(x,z)$-plane has
particular dependence on space and time, so it rotates around a center
of rotation $\R=(X,Z)$ with angular velocity $\omega$ and fiducial
phase $\Phi$
\begin{equation}
\u(\r,t)=\U(\rho (\r-\R),\vartheta (\r-\R) + \omega t - \Phi) ,
                                      \eqlabel{SW}
\end{equation}
where $\rho(\r-\R),\vartheta(\r-\R)$ are polar coordinates centered at
$\R$. A spiral wave can of course rotate in either direction; we
assume $\omega>0$ for clockwise rotation.

In presence of a small perturbation, $\epsilon\ne0$, a spiral wave
preserves the pattern, only slowly changing its frequency and location
of the core. It actually behaves as a localised object, only sensitive
to perturbations affecting its core. The localised sensitivity to
perturbations is mathematically expressed in terms of the spiral
wave's response functions, that is the critical eigenfunctions of the
adjoint linearised operator, which are essentially nonzero only in the
vicinity of the core and exponentially decay with distance from
it. Knowledge of the response functions allows quantitatively accurate
prediction of spiral waves’ drift due to small perturbations of any
nature, which makes the response functions a property that is as
fundamental for spiral waves as mass is for matter.  In particular,
the $\R$-drift velocity, \ie\ the velocity of the drift of the
position of the core of the spiral, is defined, in the first order in
$\epsilon$, by an integral of the perturbation $\h$,
\begin{equation}
\dtime{R} = \epsilon \int\limits_{\phi-\pi}^{\phi+\pi} e^{-i\xi} 
    \inner{\RF }{\tilde \h(\U;\rho,\theta,\xi) }
    \frac{\d\xi}{2\pi}+\O(\epsilon^2),  \eqlabel{forces-trans}
\end{equation}
where $R=X+\i\,Z$ is the complex coordinate of the instant spiral
centre, inner product $\inner{\cdot}{\cdot}$ stands for the scalar
product in functional space,
\[
  \inner{\mx{w}}{\mx{v}} 
  = \int\limits_{\Real^2}
  \mx{w}^+(\r) \, \mx{v}(\r) \,\d^2\r 
  = \oint \int\limits_{0}^{\infty}
  \mx{w}^+(\rho,\theta) \, \mx{v}(\rho,\theta) \rho\,\d\rho\, \d\theta,
\]
function $\tilde\h$ is perturbation $\h$ of the right-hand side in
\eq{RDS}, re-written in the $\R$-centered corotating frame of
reference $(\rho,\theta)$, where $\theta=\vartheta+\omega t-\Phi(t)$
is the polar angle in the corotating frame of reference,
and $\phi=\omega t - \Phi(t)$ is the time measured in terms of the
spiral rotational phase.  The kernel $\RF(\rho,\theta)\in\Complex$ of
this integral is the (translational) response function which
characterizes the unperturbed spiral wave solution \eq{SW} and can be
calculated numerically together with it.  Given the dependence of the
perturbation $\tilde\h$ on the current position of the spiral $R$,
equation~\eq{forces-trans} is a closed system of ordinary differential
equations (ODEs) for the coordinates of the instant centre of rotation
of the spiral wave.

\titfigure{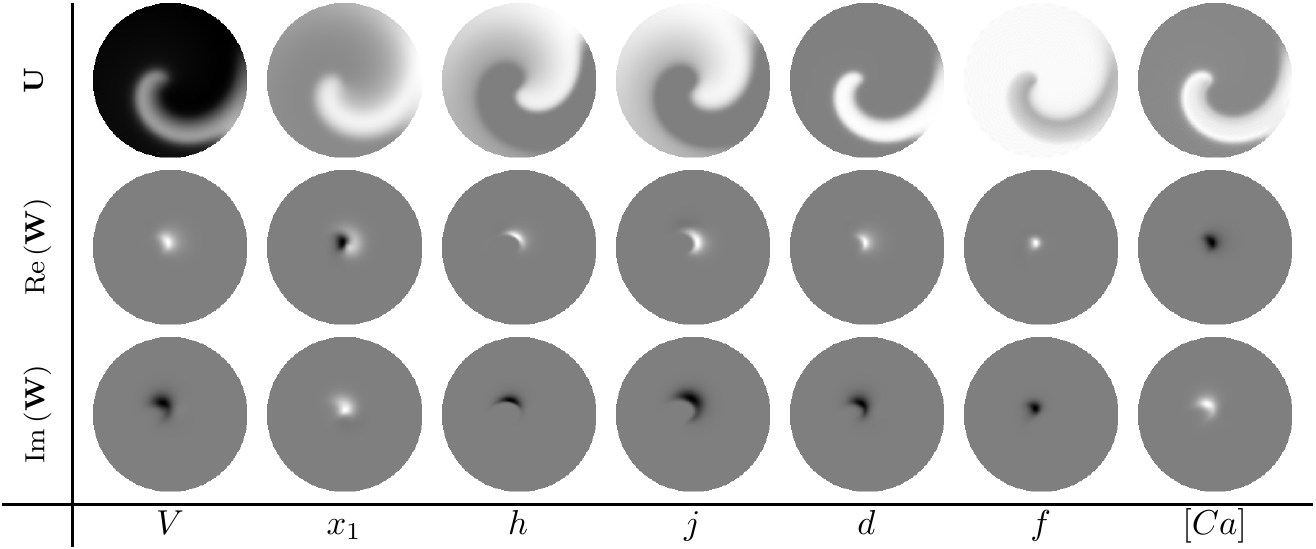}{
  Density plots of the components of a spiral wave solution {$\U$}
  and its translational response function {$\RF$}.
}{
  Parameter $\param=0.115$. The radius of the disk is $4\,\mm$ assuming
  $\Dfac=10^{-2}\,\cm^2/\s$. 
  In each plot, white corresponds to a
  value $\Ampl$
  and black corresponds to $-\Ampl$ where $\Ampl$ is chosen
  individually for each plot, \eg\ for the $V$-component of $\U$,
  $\Ampl=74.6\,\mV$. The grey periphery of the {$\RF$}
  plots, the second and
  third rows, corresponds to 0. 
}{rf}

A more detailed exposition of the theory and description of the method
of calculating the response functions are given
in~\cite{Biktasheva-etal-2009,Biktasheva-etal-2010}. In the present
study we use the same method with the modifications relevant to
the BRP model, which has $l=7$ 
as opposed to simplified $l=2$ models considered
in~\cite{Biktasheva-etal-2009,Biktasheva-etal-2010}.  \Fig{rf} shows
density plots for the spiral wave, $\U$, and its
response functions, $\RF$, in BRP model for 
$\param=0.115$; for other values of $\param$ the plots look qualitatively similar.  The
important property is that all components of the response functions are large only in
the core of the spiral and quickly decay beyond it. 

Scroll waves are three-dimensional analogues of spiral waves. They
rotate around curves called \emph{filaments}, as spiral waves rotate
around points called centres.  In general, scroll filaments are not
fixed in space but move, typically on a slow timescale relative to the
rotation period.  Hence, in addition to whatever dynamics 2D spiral
waves might have, scroll waves exhibit additional dynamics associated
with filament motion~\cite{%
  Yakushevich-1984,%
  Panfilov-Pertsov-1984,%
  Panfilov-etal-1986,%
  Panfilov-Rudenko-1987,%
  Bansagi-Steinbock-2007,%
  Luengviriya-etal-2008,%
  Luengviriya-Hauser-2008,%
  Dutta-Steinbock-2010%
}. 
Working in Frenet coordinates, the motion may be conveniently
expressed in terms of the velocities $V_N$ and $V_B$ in the normal and
binormal directions, respectively, at each point along the
filament. Motion along the tangential direction is of no physical
significance and is equivalent to reparametrization of the filament.

Then, the motion equation for the filament, in the assumption of small
filament curvature, $\curv=\O(\epsilon)$, and slowly varying phase,
has the
form~\cite{Keener-1988,Biktashev-1989,Biktashev-etal-1994,Dierckx-etal-2009}
\begin{equation}
\dtime{R} = V_N + \i V_B 
 = \senscurv \curv + \dots, \qquad \senscurv\in\Complex,   \eqlabel{fil-motion}
\end{equation}
where omitted are terms representing effects of the perturbations of
the right-hand sides, if any (which may be of the same order as that
shown), and higher-order terms.  The complex coefficient $\senscurv$
in the equation~\eq{fil-motion} is calculated using the same response
functions as for the underlying spiral wave, as
\begin{equation}
  \senscurv = -\frac12 \inner{%
    \RF (\rho,\theta)%
  }{%
    \D \e^{-\i\theta} \left(
      \@_\rho-\frac{\i}{\rho}\@_\theta
    \right) \U(\rho,\theta)
  } ,                                                       \eqlabel{senscurv}
\end{equation}
and the positive sign of $\Re{\senscurv}$ means movement towards the
local centre of curvature.

Following \cite{Keener-1988}, some publications use the notation
$\senscurv=b_2+\i c_3$. As shown in
\cite{Biktashev-1989,Biktashev-etal-1994}, the real component
$b_2=\Re{\senscurv}$ has special importance: if $b_2>0$, the overall
length of the filament becomes shorter with time, and if $b_2<0$, the
filament lengthens with time, as long as the asymptotic description
remains valid. Hence this coefficient is sometimes called
\emph{filament tension} of the scroll wave. The coefficient $c_3$ is
the {\em binormal drift coefficient} and describes the drift of a
scroll ring filament perpendicular to the plane of the ring, or more
generally, the velocity component orthogonal to the local plane of the
filament.

\paragraph{Superposition principle} 

Since the right-hand side of \eq{forces-trans} is linear in
$\epsilon\h$, the 1st-order asymptotic theory obeys a superposition
principle: if the overall perturbation is a sum of several components,
\begin{equation}
  \epsilon\h = \sum_j \epsilon_j \h_j,
\end{equation}
then the overall drift velocity is determined by the sum of the
corresponding partial ``forces'',
\begin{equation}
  \dtime{R} \approx \sum_j \sens_j \epsilon_j, 
\end{equation}
where $\epsilon_j$ is the magnitude of the $j$-th perturbation, and
$\sens_j$ is the force produced by a unit perturbation of that sort,
hereafter referred to as ``specific force'', given by
\begin{equation}
\sens_j = \oint e^{-i\xi} 
    \inner{\RF }{\tilde \h_j}
    \frac{\d\xi}{2\pi} .
\end{equation}
In the setup of our present study, the forces acting on a spiral or
scroll wave of excitation within the recovering ischaemic tissue are
caused by the filament curvature 
(described by specific force $\senscurv$),
the localised inhomogeneities and
the smooth gradient of parameter $\param$
($\sensi$ and $\sensp$ respectively),
and the gradient of diffusivity
($\sensD$).
We shall now present the explicit form of the the
relevant perturbations and the forces.
 
\paragraph{2D curvature drift.} It had been
shown~\cite{Panfilov-Pertsov-1984} that due to the axial symmetry of a
scroll ring solution, there is a strong connection between the scroll
ring filament's motion in 3D and drift of the core of a spiral wave in
response to applied electric field (electrophoretic drift) in 2D.  For
the corresponding perturbed 2D reaction-diffusion equation,
\begin{equation}
  \@_t\u = \f(\u) + \D \nabla^2 \u  + \epsr \hr,  \qquad
  \hr=\hr[\u]=\D \@_x \u ,	\eqlabel{RDSring}
\end{equation}
the specific force $\sensr$ of the electrophoretic drift is given
by~\cite{Biktasheva-etal-2010,Foulkes-etal-2010}
\begin{equation}
  \sensr = \frac12 \inner{%
    \RF (\rho,\theta)%
  }{%
    \D \e^{-\i\theta} \left(
      \@_\rho-\frac{\i}{\rho}\@_\theta
    \right) \U(\rho,\theta)
  } ,                                                       \eqlabel{sensr}
\end{equation}
which is exactly the opposite of $\senscurv$ given by \eq{senscurv}.
The opposite sign can be understood if one remembers that the positive
sign of $\Re{\senscurv}$ means movement towards the local centre of
curvature of the filament, and the form of the
perturbation~\eq{RDSring} with positive $\epsr$ corresponds to the
centre of curvature located at the line $(-1/\epsr,0)$, \ie\ in the negative
$x$ direction with respect to the current spiral
centre~\cite{Panfilov-Pertsov-1984}.

This equivalence of $\senscurv$ and $\sensr$ up to the sign allows us
to use the 2D simulations of system \eq{RDSring} to estimate the drift
velocity
of a 3D scroll ring, and hence estimate the 3D coefficient
$\senscurv=-\sensr$. Subsequently, these 2D estimations can be used to
verify/confirm both drift velocities $\sensr$ and $\senscurv$ obtained
using the response functions in \eq{sensr} and \eq{senscurv}.

\paragraph{Smooth gradient of excitability.}
We suppose that the excitability kinetic parameter $\param$ varies in space,
\begin{equation}
  \f = \f(\u,\param), \qquad \param = \param(\r), 
\end{equation}
and, further, that the profile $\param(\r)$ is smooth enough and can
be approximated by a linear spatial gradient, within the spiral core
where the components of the response functions are essentially
non-zero,
\begin{equation}
  \param(\r) \approx \param_0 + \vec{\epsp} \cdot (\r-\R), 
  \qquad
  \param_0=\param(\R), 
  \quad 
  \vec{\epsp} = \nabla\left.\param(\r)\right|_{\r=\R}. 
\end{equation}
Then, the velocity of the drift induced by the parameter $\param$
gradient works out~\cite{Biktasheva-etal-2010} as
\begin{eqnarray}
 && \dtime{R} = \sensp \epsp,  \nonumber\\
 && \sensp=\frac12 \inner{\RF}{e^{-\i\theta} \dfdp(\U;\param_0)}, \nonumber\\
 && \epsp=\left( \@_x + \i\@_z \right) \left.p(\r)\right|_{\r=\R}. 
\end{eqnarray}
The real part of $\sensp$ gives the component of the drift velocity
along the gradient of $\param$ and is positive if the drift is towards
higher values of $\param$. The imaginary part of $\sensp$ describes
the drift across the gradient of $\param$; it is positive if the
lateral component of the drift velocity is counter-clockwise with
respect to the direction of $\nabla\param$.

\paragraph{Localized inhomogeneity of excitability.}
As can be seen from \fig{rf}, the core size of the spiral wave in BRP
model is $\sim1\mm$ for $\Dfac=10^{-2}\,\cm^2/\sec$. A 1000-fold
decrease of $\Dfac$ down to $10^{-5}\,\cm^2/\sec$ implies shrinkage of
the core to the size of one cell, $\sim30\,\um$. Hence for the
coupling values at the lower end of the range, localized
heterogeneities of $\param$ become of principal importance, and they
cannot be considered as smooth gradients.

To elucidate possible role of the localised inhomogeneities, let us
consider the case when the continuous limit is still applicable, but
the spiral core size is comparable with the size of a localized
inhomogeneity, or the magnitude of such inhomogeneity is so
significant it affects the spiral dynamics despite the small geometry
size. This can happen when the random distribution of properties
produces relatively large lumps of cells with local average
excitability deviating from the overall average.  Let's consider an
idealized situation when the parametric inhomogeneity is localized in
a disk of radius $\Ri$ centered at $\Rc=(\Xc,\Zc)$ and is uniform
within it, so
\begin{equation}
  \param(\r)=\param_0+\epsi\param_1(\r),
  \qquad
  \param_1=\frac{1}{\pi\Ri^2}\,\Heav(\Ri-|\r-\Rc|) , 
\end{equation}
where $\Heav(x)$ is the Heaviside step function.  Then for a small
enough $\Ri$, the velocity of the drift induced by the localized
inhomogeneity is defined
as~\cite{Biktashev-etal-2010,Biktasheva-etal-2010}
\begin{equation}
  \dtime{R} = \sensi \epsi, \qquad 
  \sensi = - \frac{R-\rc}{|R-\rc|} \F(|R-\rc|) , \label{EoM}
\end{equation}
where  $\rc=\Xc+\i\Zc$ and
\begin{equation}
  \F(\distd) = \oint \e^{-\i\theta}\,
  \left[\RF (\distd,\theta)\right]^+ \, \@_{\param}\f(\U(\distd,\theta);\param_0) 
  \,\frac{\d\theta}{2\pi} + \O(\Ri).
                                        \label{central-force}
\end{equation}
Here $\Re{\F}$ is the radial component of the drift velocity, positive
if the spiral moves towards the centre of the inhomogeneity, and
$\Im{\F}$ is its azimuthal component, positive if clockwise with
respect to the centre of inhomogeneity.

\paragraph{Gradient of the diffusivity}
We also deal with the drift caused by a gradient of the diffusivity,
so that
\begin{equation}
  \@_t\u=\n \nabla\left( \Dfac(\r) \nabla \u \right) +
  \f(\u)                                                    \eqlabel{vardiff}
\end{equation}
Suppose the diffusivity varies smoothly, so it can be approximated by a
linear function within the core of the spiral,
\begin{equation}
  \Dfac(\r) \approx \Dfac_0  + \vec{\epsD} \cdot (\r-\R), 
  \qquad
  \Dfac_0=\Dfac(\R), 
  \quad 
  \vec{\epsD} = \nabla\left.\Dfac(\r)\right|_{\r=\R}. 
\end{equation}
Substituting this into \eq{vardiff}, we get the perturbed
reaction-diffusion equation of the form \eq{RDS} with $\Dfac=\Dfac_0$
and the perturbation
\begin{equation}
  \epsD \hD = \Dfac_0 (\vec\epsD\cdot\nabla) \n\u 
  + (\epsD\cdot(\r-\R)) \Dfac\nabla^2 \n\u .        \eqlabel{hdiff}
\end{equation}
This leads to the expression for the specific force induced by the
gradient of the diffusivity in the form
\begin{equation}
  \sensD = \sensD^{(1)} + \sensD^{(2)}                      \eqlabel{sensD}
\end{equation}
where
\begin{equation}
  \sensD^{(1)} = \frac12 \inner{%
    \RF (\rho,\theta)%
  }{%
    \n\Dfac_0 \e^{-\i\theta} \left(
      \@_\rho-\frac{\i}{\rho}\@_\theta
    \right) \U(\rho,\theta)
  } ,                                                       \eqlabel{sensD1}
\end{equation}
and
\begin{equation}
  \sensD^{(2)} = \frac12 \inner{\RF (\rho,\theta)}{
   \rho\,\e^{-\i\theta} 
   \,\n\Dfac_0\nabla^2\U(\rho,\theta;\param_0)
 }.                                                         \eqlabel{sensD2}
\end{equation}
It is easy to see that the specific force $\sensD^{(1)}$ in
  \eq{sensD1} coincides with the 2D electrophoretic drift specific
  force $\sensr$ given by \eq{sensr} up to the substution
  $\Dfac=\Dfac_0$. On the other hand, Dierckx
  ~\cite{Dierckx-etal-2009} has shown that the problem of drift in the
  gradient of diffusivity is equivalent to the problem of 2D
  electrophoretic drift, up to a transformation of coordinates. This
  implies that $\sensD=\sensr$, and since $\sensD^{(1)}=\sensr$, the
  integral \eq{sensD2} should be zero.  In our calculations 
  using response functions, the values of
  $|\sensD^{(2)}|$ do not exceed $3\times10^{-6}$ for the whole range
  of $\param_0$ considered, which is indeed small compared with
  typical values of $|\sensD|$ shown
  in~\fig{alp-graphs}(a) (note that $|\sensD|=|\senscurv|$). This
  small deviation of the calculated value of $\sensD^{(2)}$ from zero
  serves as a measure of accuracy of the response function and
  the integrals based on it. 
 Note that specific forces correspond to the limit of
  $\epsilon_j\to0$.  Direct numerical simulations presented
  in~\cite{Sridhar-etal-2010}, performed at finite values of $\epsD$
  and in a different model, show empirical values of $\sensD$ and
  $\sensr$ differing by as much as $10\%$.

\section*{Results}

\subsection*{Continuous limit: predictions from the asymptotic theory}

\subsubsection*{Effects of elementary perturbations} 

\titfigure{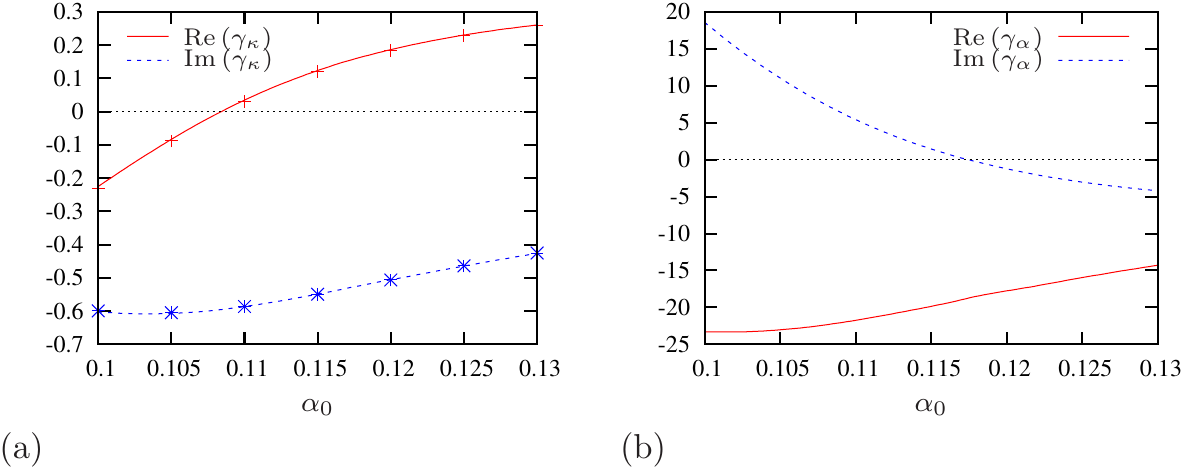}{
  Dependence of the specific forces $\senscurv$ and $\sensp$
  on the unperturbed value of
  the excitability parameter $\param_0$.
}{Diffusion coefficient $\Dfac=10^{-2}\,\cm^2/\s$.
  (a) Specific force $\senscurv$
  caused by filament curvature $\curv$.
  Note that $\senscurv=-\sensr=-\sensD$.
  Symbols ``+'' and ``$\ast$'' show estimates of $-\sensr$ from
  direct numerical simulations of~\eq{RDSring}, for comparison. The
  difference between predicted and simulation values is smaller than
  $0.005$ at all points. 
  (b) Specific force $\sensp$ caused by the gradient of excitability
  parameter $\param$. 
  Red solid lines: real parts, the longitudinal components. 
  Dashed blue lines: imaginary parts, the lateral components. 
  The meanings of the vertical axes are different for
  different curves and are designated in the legends.
}{alp-graphs}

Based on the response functions shown in \Fig{rf}, we have computed
the values of the specific forces acting on spiral (2D) and scroll
(3D) waves under conditions associated with recovering ischaemic
border. These forces include the specific force 
$\senscurv$ 
caused by the
curvature of the vortex filament, the specific force 
$\sensD$ 
caused by the
gradient of diffusivity, the specific force 
$\sensp$ 
caused by the gradient of
parameter $\param$ and the specific force 
$\sensi$ 
caused by a localised
inhomogeneity of parameter $\param$.

\Fig{alp-graphs}(a) shows the theoretical predictions for the
components of the specific force $\senscurv$ caused by the curvature
of the vortex filament. 
That panel also shows an excellent agreement of 
  these predictions
  with the results of the direct numerical simulations of electrophoretic
  drift~\eq{RDSring} (remember that $\senscurv=-\sensr$).
The components of $\senscurv$ correspond to
the two filament's drift coefficients: the ``filament tension''
$b_2=\Re{\senscurv}$, and the {\em binormal drift coefficient}
$c_3=\Im{\senscurv}$. The {\em binormal drift coefficient} $c_3$
determines \eg the drift of scroll rings along their axis. The
filament tension $b_2$ is usually much more important for a scroll's
dynamic, as the positive filament tension means that the filament will
tend to straighten or collapse if geometry allows it. Negative
filament tension means that the filament will tend to spontaneously
lengthen and curve and can produce ``scroll wave turbulence'' which is
phenomenologically similar to fibrillation~\cite{%
  Panfilov-Pertsov-1984,%
  Brazhnik-etal-1987,%
  Biktashev-1989,%
  Biktashev-etal-1994,%
  Biktashev-1998%
}.  An important observation from \Fig{alp-graphs}(a) is that in the
shown interval of parameter $\param$, filament tension 
$b_2=\Re{\senscurv}$
changes the
sign and is overall smaller than the binormal drift coefficient $c_3$.

As discussed above, the drift caused by the curvature of the filament
is equivalent to the drift caused by the gradient of the diffusion
coefficient, so the same coefficients, though taken with the opposite
sign, will describe the drift of the spiral core or scroll filament in
response to gradient of diffusivity.  Namely, coefficient
$\Re{\sensD}=-\Re{\senscurv}=-b_2$
will determine the component of the drift along the gradient of diffusivity and 
$\Im{\sensD}=-\Im{\senscurv}=-c_3$
across it.  Following \fig{alp-graphs}(a),
$\Re{\sensD}=-\Re{\senscurv}<0$ at higher values of $\param$, and
$\Re{\sensD}=-\Re{\senscurv}>0$ at lower values of $\param$.  So, at
higher values of $\param$ the negative specific force of the gradient
of diffusivity will drag the spirals towards poor coupled regions with
smaller diffusion, while at lower values of $\param$ the positive
specific force of the gradient of diffusivity will drag the spirals
towards better coupled regions with higher diffusion. Thus, the fact
that $b_2$ changes sign in the relevant range of parameters, means
that the diffusivity gradient can either drag spirals towards the
pourly coupled bottom layer or repell them into the better coupled
upper layer, depending on the local value of excitability parameter
$\param$. Also, the fact that $|b_2|<|c_3|$ means that the spirals
should move preferentially across the diffusivity gradient that is
along the border zone, which agrees with the numerics and experiments.

\Fig{alp-graphs}(b) shows the theoretical predictions for the drift
coefficients in response to a smooth gradient of parameter
$\param$. Here, an important feature is that the longitudinal
coefficient $\Re{\sensp}$ is negative in the whole range of
$\param_0$. This means that the spirals should drift towards areas
with lower excitability. This agrees with the general
  rule noted \eg\ in~\cite{Rudenko-Panfilov-1983,TenTussher-Panfilov-2003}.

\titfigure{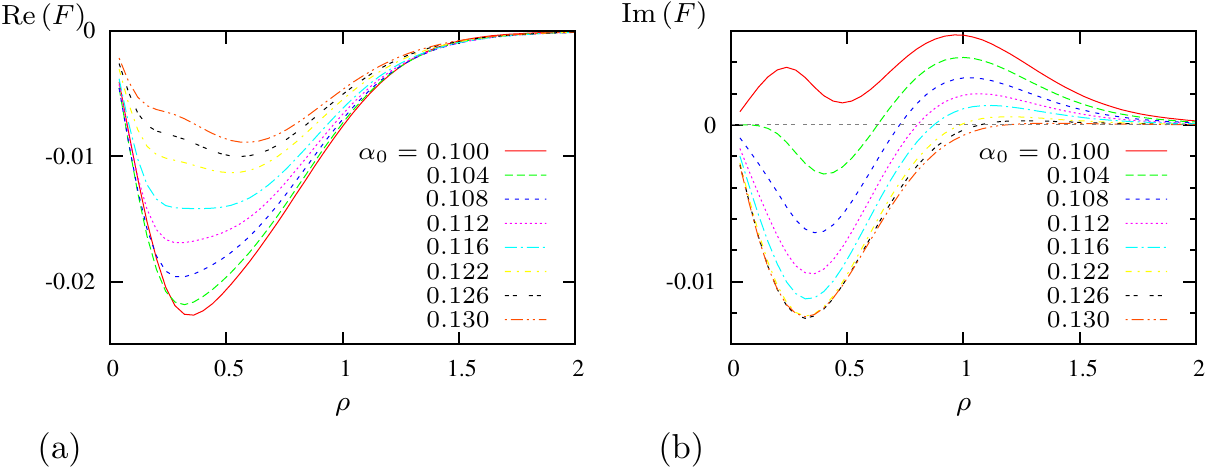}{
  Interaction with point-like inhomogeneity.
}{
  Dependences of (a) radial and (b) tangential components of the
  specific force caused by point-like inhomogeneity of excitability $\param$, 
  on the distance from instant spiral rotation centre 
  to the inhomogeneity, at selected values of background excitability $\param_0$
  as indicated in the legends. 
  The scale of $\distd$ is given in $\mm$ assuming $\Dfac=10^{-2}\,\cm^2/\s$. 
}{pin}

\Fig{pin} shows the theoretical prediction for interaction of a spiral
wave with a point-like heterogeneity in parameter $\param$. Here, the
interaction force depends on the distance between the spiral's centre
and the heterogeneity. The negative sign of the radial component
$\Re{\F(\distd)}$, observed for all distances $\distd$ and all values
of $\param$ considered, means that a localized inhomogeneity with
lowered excitability, $\epsi<0$, should attract spiral waves, and
those with higher than the background excitability, $\epsi>0$ should
repel them. This is also intuitively consistent with the predictions
for the linear gradient of $\param$ given by \fig{alp-graphs}(b).

The constant sign of the inhomogeneity specific force $\sensi$ radial
component $\Re{\F(\distd)}$ in \fig{pin}(a) is not a general case, and
in other models the sign of interaction with a localized inhomogeneity
may depend on the distance to it, which may lead to ``orbital'' motion
around such inhomogeneity, with orbit radii at the zeros of
$\Re{\F(\distd)}$~\cite{Biktashev-etal-2010}. So, following the graphs
in \fig{pin}(a) and the shown constant sign of the radial component
$\Re{\F(\distd)}$, we should not observe an orbital motion in our
present BRP model.

\subsubsection*{Complex perturbations and pinning/unpinning in 2D} 

We shall now use the superposition principle to analyse the 2D drift
of a spiral wave subject to a combination of forces caused by a smooth
gradient of diffusivity, a smooth gradient of the excitability
parameter $\param$ and a localised inhomogeneity in parameter
$\param$.  In a system of reference with the origin at the centre of
the disk inhomogeneity, $\Rc=\vec{0}$, the equation of motion for a
complex coordinate of spiral wave rotation centre $R$ is
\begin{equation}
\dtime{R} = - \epsi \frac{R}{|R|} \F(|R|) + \Grad     \eqlabel{pin-grad}
\end{equation}
where 
$\F$ is the force induced by the localised inhomogeneity and
$\Grad$ is the constant dragging force due to the smooth
gradient of parameter $\param$ and/or the diffusivity gradient.  We
use polar coordinates for the instant centre position, $R = \distd
\e^{\i\angd}$, and also set $\Grad=\Gradm\e^{\i\Gradang}$ where
$\Gradm$ and $\Gradang$ are the magnitude and direction of the
gradient force. Further, we separate the radial and azimuthal
components of force $\F$, 
$\F(\distd)=\Fc(\distd)+\i\Fa(\distd)$. Then, the equations of motion
in the two real dynamic variables are
\begin{eqnarray}
  \dtime{\distd} &=& -\epsi\Fc(\distd) + \Gradm \cos(\Gradang-\angd), \nonumber\\
  \distd\,\dtime{\angd} &=& -\epsi\Fa(\distd) + \Gradm \sin(\Gradang-\angd) . \eqlabel{pin-grad-2}
\end{eqnarray}
An equilibrium in the system \eq{pin-grad-2} may be observed at a
radius $\distd$ satisfying
\begin{equation}
  \epsi^2(\Fc^2(\distd) + \Fa^2(\distd)) = \Gradm^2. \eqlabel{equil}
\end{equation}
It is easy to see that equilibria will not exist, that is, the smooth
gradient force will definitely tear a spiral off from the localized
inhomogeneity, if
\begin{equation}
  |\Gradm|>\Gradm_{\crit}
  = |\epsi| \max\limits_{\distd}\left( \Fc^2(\distd) + \Fa^2(\distd)\right)^{1/2}
  = |\epsi| \max\limits_{\distd}\left( |\F(\distd)| \right),
                                                  \eqlabel{tear-thresh}
\end{equation}
that is, if the gradient force exceeds the maximal force of
interaction with inhomogeneity, including both radial and azimuthal
components (see also \cite{Pazo-etal-2004} where a special case with
$\Fa(\distd)\equiv0$ was considered).

\titfigure{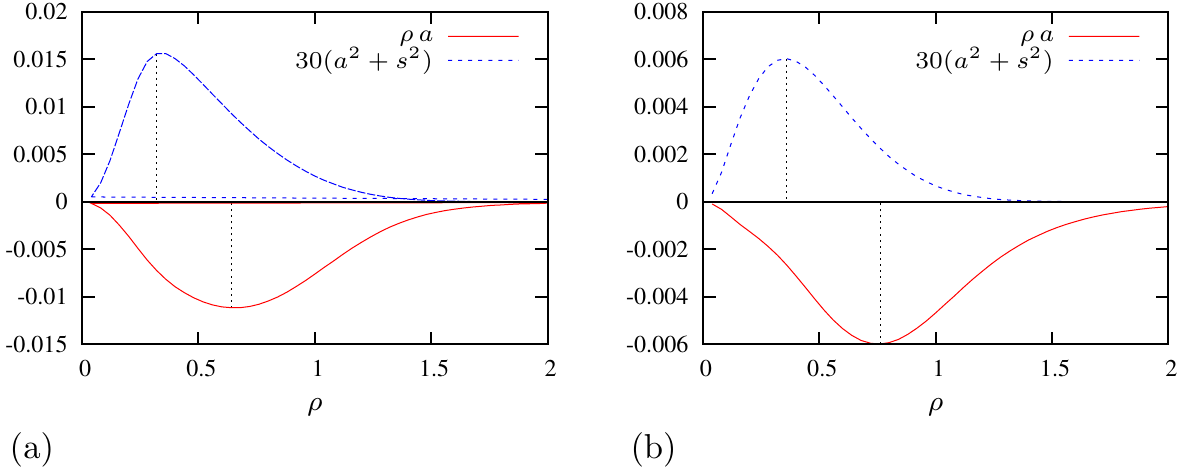}{
  Graphs for graphical solution of stability of ``pinning
  equilibrium''. 
}{
  (a) $\param_0=0.10$, (b) $\param_0=0.13$. 
  Diffusivity is assumed $\Dfac=10^{-2}\,\cm^2/\s$. 
  The meanings of the vertical axes are different for
  different curves and are designated in the legends.
}{pin-g}

Following \eq{equil}, for every $|\Gradm|<\Gradm_{\crit}$ there are at
least two equilibria at different values of $\distd$. Note that 
$|\Gradm|<\Gradm_{\crit}$  can happen
at either sign of $\Fc$, \ie\ both for attracting and
repelling inhomogeneities.

Standard calculations give that an equilibrium at a distance
$\distdeq$ from the inhomogeneity will be stable in linear
approximation if and only if the following two conditions are
satisfied simultaneously:
\begin{eqnarray}
 && \left.\Df{}{\distd}\left(
     \Fc^2 (\distd) + \Fa^2(\distd) 
   \right)\right|_{\distd=\distdeq} > 0 ,\nonumber\\
 && \left.\Df{}{\distd}\left(
     \epsi \distd \Fc(\distd) 
   \right)\right|_{\distd=\distdeq} > 0.  \eqlabel{Stability}
\end{eqnarray}
The stability  conditions \eq{Stability}
can be easily checked graphically, and the graphs of
the two functions involved are shown in \fig{pin-g}.  These
conditions require that both functions should
be increasing at $\distd=\distdeq$.
The first inequality does not
depend on the sign of $\epsi$, and it therefore demands that
$\distdeq$ is smaller than the position of the maximum of
$\Fc^2(\distd)+\Fa^2(\distd)$ (the blue dashed curves).  
For the second
inequality the situation is more complicated as it depends on the sign
of $\epsi$.  For the case $\epsi>0$, \ie\ repelling inhomogeneity with
excitability higher than $\param_0$, the 
second
stability condition demands
that the position of the equilibrium is to the right of the minimum of
$\distd\Fc(\distd)$, which is shown by the solid red curve.  For both
values of $\param_0$ shown in \fig{pin-g}, and also for all $\param_0$
in between, as we have checked, this is incompatible with the first
condition, as the red minimum always happens to the right of the blue
maximum.  For $\epsi<0$, \ie\ attracting inhomogeneity with
excitability lower than $\param_0$ around it, the 
second
stability condition
demands that the position of the equilibrium should 
be to the left of the red minimum, 
which is a requirement that is weaker than
the first condition, as
all points to the left of the blue maximum are also to the left of the
red minimum.  So, in our model there cannot be a stable equilibrium
near a repelling inhomogeneity, but only near an attractive
inhomogeneity.  Intriguingly, if the relative position of the two
extrema was different, \ie\ the red minimum was to the left of the
blue maximum, it would create a paradoxical possibility of a stable
equilibrium occuring due to interaction with a repelling
inhomogeneity. We are not aware of any reasons why this could not
happen in some models, but it does not happen in our present model in
the range of parameters that we are interested in.

\titfigure{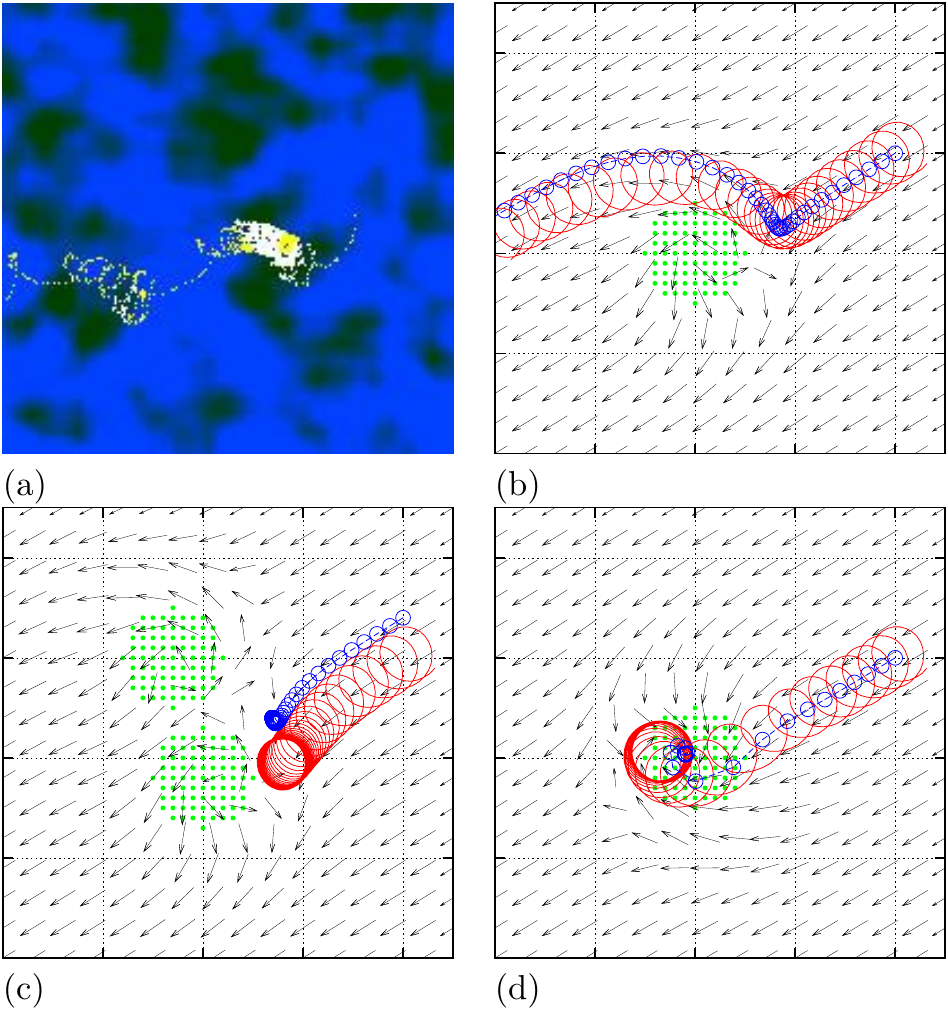}{
  Pinning of spiral wave's drift to localized inhomogeneities. 
}{
  (a) An extension of the drift trajectory shown in figure~8D
  in~\cite{Biktashev-etal-2008} with temporary pinning to a
  high-$\param$ cluster. This is a $25\times25$-cell fragment of a tip
  path in a simulation in a box of $100\times100$ cells,
  $\alpavg=0.12$,
  $\Dmin=5\cdot10^{-5}\,\cm^2/\sec$,
  $\Dmax=2\cdot10^{-3}\,\cm^2/\sec$, 
  $\c=1/6\,\cell/\sec$. 
  The colour background shows distribution of $\noise(x,y)$,
  smoothened by sliding averaging, (greenish) dark corresponds to 
  high $\param$ and (blue) light corresponds to low $\param$. 
  (b) 
  Drift caused by a repelling circular inhomogeneity
  (green dots show affected cells) of radius of $\Ri=5$~cells. Red
  solid line is the tip trajectory in 
a $100\times100$-cell simulation,
  $\param=0.13$ in the bulk of the medium and $\param=0.15$ within the
  disk, and diffusivity $\Dfac=10^{-3}\exp(\beta
  (y-y_0))\,\cm^2/\sec$, where $\beta=0.7\,\mm^{-1}$ and $y_0$ is the
  middle of the box.
  The arrows represent the corresponding direction field in the ODE
  model~\eq{pin-grad}.
  The small blue open circles are
   the instantaneous centres of rotation of the spiral predicted by
  the ODE model and shown at intervals corresponding to one rotation
  period of the spiral. These instantaneous centres of rotation of the spiral
 can be thought
  of as  sliding 
  period-averaged positions of the tip, and make a drift trajectory as predicted by
  the ODE model.
  %
  (c) 
  Two repelling inhomogeneities of the same kind as in (b) can stop
  the drift altogether.
  (d)
  Attractive inhomogeneity with lowered excitability, $\param=0.11$,
  within the disk of the same size as in (b). Now the spiral is
  permanently stalled behind the heterogeneity.  Here and elsewhere,
  the
  tip of the numerical spiral at any given moment of time is defined as an
  intersection of isolines $\V=-35\,\mV$ and $\fgate=0.85$ ($\fgate$
  is the dimensionless inactivation gating variable for the slow
  inward current).
}{pin-unpin}

Thus, these theoretical predictions based on the response functions of
the vortices suggest that stable pinning of a spiral wave in our model
may be to lowered-excitability sites only, while in the experimental
and simulations described in \cite{Biktashev-etal-2008} the pinning to
inhomogeneities of either sign was observed.  We have a closer look at
this seeming contradiction below.

Firstly, the pinning observed in experiments and simulations was not
permanent but temporary. An explanation for that could be that the
pinning persisted only until the gradient force exceeded the tear-off
threshold \eq{tear-thresh}. However, it is also possible that the
pinning was temporary because it was really a slow-down near an
unstable equilibrium in the vicinity of a repelling inhomogeneity.
Panel (a) in \fig{pin-unpin} reproduces the tip trajectory in a
`pinning'' event from \cite{Biktashev-etal-2008}, revealing that it
was actually only a temporary stall between two fast-drift
episodes. Panel (b) illustrates that this sort of stalling is easily
reproduced in deliberately arranged simulations and is well described
by the ODE model~\eq{pin-grad}.

Secondly, a certain mutual allocation of repelling heterogeneities may
cause `permanent' pinning, again until the parameters change. This is
illustrated in \fig{pin-unpin}(c). There are two identical repelling
inhomogeneities. For the given initial position of the spiral wave, if
only the lower inhomogeneity was present, the drift would proceed
along a trajectory similar to that in panel (b). However this drift is
disallowed by the presence of the upper repelling inhomogeneity, hence the
spiral stops at a point of equilibrium of three forces: the constant
dragging force and the two repulsion forces from the two localized
repelling inhomogeneities.

Panel (d) in \fig{pin-unpin} is given for completeness, to illustrate
the more straightforward case of pinning in the vicinity of an
attracting inhomogeneity. It is worth noticing a simple
phenomenological difference between pinning to a repulsive
inhomogeneity and to an attractive one: for the former, the spiral
wave stops \emph{in front} of the inhomogeneity, and for the latter,
\emph{behind} it.

There are several factors responsible for the quantitative discrepancies
seen in 
\fig{pin-unpin}(b--d) between the theoretical predictions for the trajectories
of the spiral drift and the trajectories obtained from direct
numerical simulations: 
large value of $\epsi$ affecting
the applicability of the asymptotic theory, the crudeness of the cell
structure affecting the behaviour of the direct simulations as
compared to the continuous limit, and also the finite $\Ri$ used in
simulations as compared to the small-$\Ri$ limit assumed in the
theory. 
However, the theoretical trajectories
and those obtained from direct numerical simulations
are in good
qualitative agreement, so the asymptotic theory works really well for
this complicated arrangement, despite all the simplifications made.

Naturally, with the random distribution of heterogeneity, as present
in the experiments and the numerical simulations of the ischaemic
border zone, all of the above scenarios with pinning to
inhomogeneities of either sign could take place from time to time. In
some experiments the pinning locations subsequently became sources of
ectopic waves of excitation and therefore were associated with points
of higher excitability. In other experiments, the pinning locations
never produced the ectopic waves, which suggested that the pinning
inhomogeneity had the lowered excitability. Understanding that there
are different mechanisms of pinning to attractive inhomogeneity with
lowered excitability and to a (group of) repelling inhomogeneity(s)
with elevated excitability provides an explanation for these seemingly
contradicting experimental observations.

\subsection*{Generation of a 3D turbulent pattern}

\titfigure{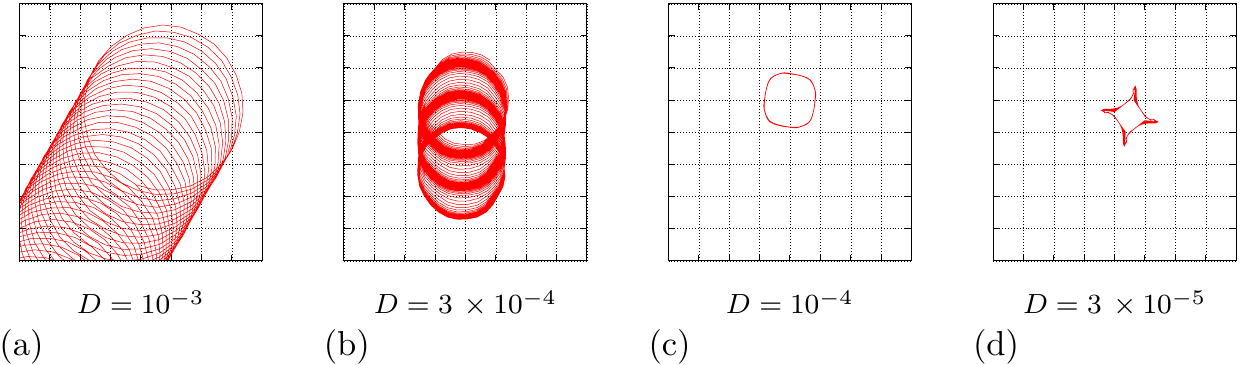}{
 Two-dimensional curvature drift.
}{
  $\param=0.13$, $\epsr=0.1\,\mm^{-1}$. Shown are tip trajectories in system~\eq{RDSring} in
  $100\times100$-cell box for various $\Dfac$, as shown under the panels, in $\cm^2/\sec$.
  Smaller diffusivity means stronger effect of the discreteness of the
  tissue, which can stop the drift altogether
  (the grid of dotted lines designates individual cells). 
}{ft}


The asymptotic theory of spiral and scroll drift is valid for PDEs,
describing continuous media. 
The theory might not be applicable if the discreteness of the cell
structure is significant when the diffusivity is small.
Thus our findings
here are purely empirical, based on direct numerical simulations.
The role of discreteness in 2D dynamics was
extensively analysed in \cite{Pumir-etal-2005,Biktashev-etal-2008} so
here we concentrate on 3D aspects. 

The effect of dicreteness is to a certain extent similar to that of
heterogeneities, \ie\ it can hinder the drift caused by the smooth
parametric or diffusivity gradient. This is illustrated in \fig{ft},
where we present simulations of 2D curvature-induced 
(electrophoretic)
drift, for
different values of diffusivity at $\param=0.13$ and positive
$\epsr$. As can be seen from \Fig{alp-graphs}(a), at $\param=0.13$
$\Re{\senscurv}>0$, so $\Re{\sensr}=-\Re{\senscurv}$ is negative which
corresponds to the drift in the negative $x$ direction in \fig{ft}.

It can be seen that in \fig{ft}, in line with continuous limit
predictions, as diffusivity decreases, so does the spatial scale of
the spiral tip trajectory. Further still there is another effect which
has an entirely discrete nature: as diffusivity becomes too small, the
drift of the spiral stops altogether. Panel (b) indicates that there
is a range of diffusivities at which the longitudinal component of the
drift (which corresponds to the filament tension $b_2$ and is smaller
in absolute value than the lateral component, see
\Fig{alp-graphs}(a)), `freezes out', while the lateral component is
still observed, so the drift proceeds along the vertical grid line.

Note that change of filament tension due to relatively small
  discreteness is a generic feature of excitable media, and has been
  reported in FitzHugh-Nagumo~\cite{Foulkes-etal-2010} and
  Barkley~\cite{Alonso-etal-2011} models.

\titfigure{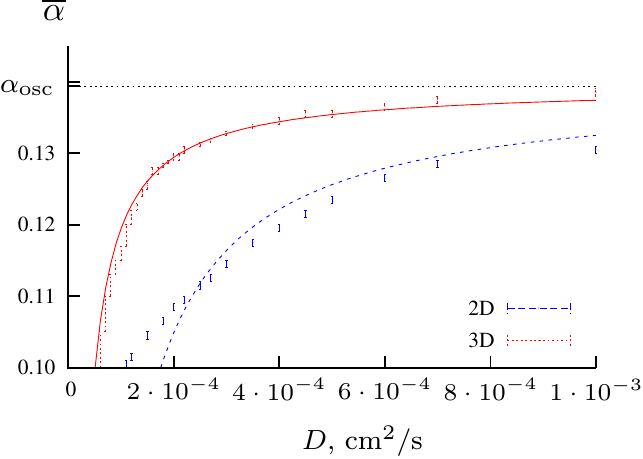}{
  The transition curves obtained in simulations of a 2D and a thin 3D
  layer of cells. 
}{
  $\delpar=0.5$.
  Below each corresponding curve, the system is quiescent, above the
  curve and below the $\alpavg=\alposc$ line, focal sources are
  observed.  We also show the best fits, with the weight
  $\propto\Dfac^2$, by the theoretical dependence
  $\alpavg\approx\alposc-B^2/D$ suggested in \cite{Pumir-etal-2005} .
}{parport}


Another important effect of the tissue discreteness is due to the role
of microscopic heterogeneities of parameter $\param$, defined by
equation~\eq{heterogeneity}, in the generation of ectopic foci and
breakup of excitation waves. In presence of the microscopic
heterogeneity, $\delpar>0$, a macroscopically homogeneous tissue, with
$\Dfac=\Dmin=\Dmax$, $z_1=-\infty$ in \eq{Dzt} and \eq{alpzt}, may
either show spontaneous focal sources or be quiescent depending on
particular combination of $\Dfac$, $\paravg$ and $\delpar$. The
critical curves in the $(\Dfac,\paravg)$ plane, separating the
automatic and exctiable regimes (corresponding to the zones III and V
in~\fig{diagram}
and transition between them), are shown in
\fig{parport}, for 2D and 3D cases. We obtained the 3D curve from
direct simulations on a thin three-dimensional grid of
$40\times60\times5$ cells at $\delpar=0.5$. The transition curve
obtained from 2D simulations, as in \cite{Pumir-etal-2005}, is shown
on the same graph for comparison. One can see that position of the 3D
transitional curve is elevated compared to the 2D transitional
curve. This elevation is due to the fact that every cell in 3D is
connected to more neighbours, which increases the load on the
automatic cells surrounded by non-automatic environment. Therefore, in
3D it takes more automaticity to overcome the coupling with the
quiescent neighbours, so in 3D simulations the same regimes are
observed at different values of parameters $\paravg$ and $\Dfac$ than
in 2D simulations.

We performed numerical simulations of the 3D tissue slab with
diffusivity and excitability profiles shown in~\fig{bubbles}(c). The
lower layer contained fully uncoupled cells with excitability
$\param=0$. This layer corresponded to the region IV
in~\fig{diagram}, ``a
quiescent state where wave propagation is not possible''.
To reveal the above described effects of
tissue discreteness, we performed simulations using three different sets of
parameters $\Dfac$, $\paravg$ and $\delpar$ defining properties of the
upper layer. We used $\Dmin=\Dmax/100$ in all cases.

The first, ``toy'' set of parameters (\figs{bubbles} and
\figref{scrolls}) was $\Dmax=5\cdot10^{-5}\,\cm^2/\s$, $\parmax=0.13$,
$\delpar=0.1$. At the small $\delpar=0.1$ the number of cells getting
above $\alpavg=\alposc$ line will be small, resulting in further
elevation of the 3D transitional curve compared to the $\delpar=0.5$
shown in \fig{parport}. 
Therefore,
the upper layer with this set of
parameters $\Dfac$, $\paravg$ and $\delpar$ still corresponded to the
region V ``a quiescent state where wave propagation is possible'',
which ensured a
transition 
from what is
described as region III ``fragmented
ectopic waves '' within the middle layer 
to region V in the top layer. The value
$\Dmin\propto10^{-7}\,\cm^2/\s$ was below the physiologically
meaningful range, so simulations with the smaller $\Dfac$ were more of
a mathematical excercise, which allowed however, due to the smaller
spatial scales involed, to perform a relatively detailed study despite
the computational expences of the three-dimensional model. The
principal conclusion was then tested with the more physiologically
relevant set of paramers.

The two ``more realistic'' sets of parameters (\fig{large-D}) were
$\Dmax=10^{-3}\,\cm^2/s$, $\delpar=0.5$, with either $\parmax=0.105$ or
$\parmax=0.115$, both corresponded to the region V
``a quiescent state
where wave propagation is possible''
(below the critical line in~\fig{parport}), 
which also ensured the transition from region III to
region V within the middle layer. This sets of parameters were ``more
realistic'' in terms of the value of diffusivity more relevant to
physiologically meaningful range.

\titfigure{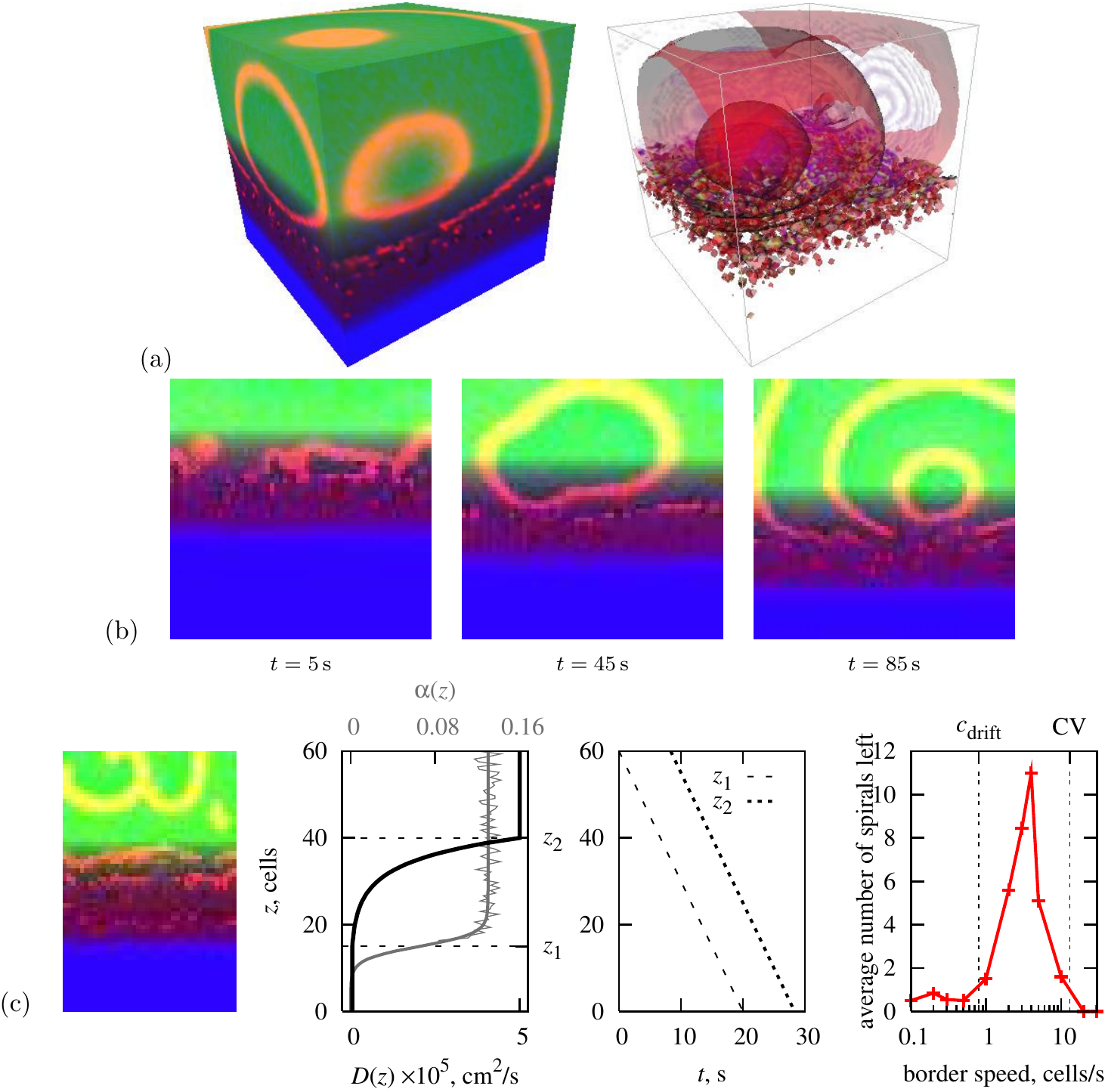}{
  The ischaemic border zone in three dimensions.
}{
  ``Toy'' set of parameters: $\Dmax=5\cdot10^{-5}\,\cm^2/\s$,
  $\parmax=0.13$, $\delpar=0.1$.
  (a,b) Box size
  $60\times60\times60$ cells,
  border speed $\c=1/6\,\cell/\s$.
  (a) Snapshot of activity on the surface and inside the box (red
  semi-transparent surfaces are excitation fronts).
  (b) Activation patterns on a middle cross-section of the box.  
  (c) Schematic of the study of spirals' probability to escape to the
  well coupled zone: a snapshot through the middle of a thin 3D layer
  of cells (box size $40\times5\times60$, border speed $\c=3\,\cell/\s$);
  corresponding distribution of $D$, $\paravg$ and $\param$;
  movement of boundaries with time; and average number of spirals left
  in the box after passing of the border zone, as function of the its
  speed. Here $c_{\mathrm{drift}}$ is a typical drift velocity and CV
  is a typical conduction velocity.  
}{bubbles}

\Fig{bubbles} presents a simulation with the ``toy'' set of
parameters in a box size $40\times30\times60$ cells and a relatively
slow border speed of $\c=1/6\,\cell/\s$. 
Panel (a) shows small ectopic sources giving rise to multiple
ectopic ``bubbles'' in 3D, also shown in cross-sections from the
model's cube of cells in \fig{bubbles}(b).  The wavefronts from
multiple smaller ectopic sources fused into larger wavefronts which
were spreading toward the upper, better coupled layers of the border
zone.
No scroll wave activity is observed in the upper zone. 
When the transitional border zone has passed down, the cube is
left without ectopics and all activity is ceased. 

\Fig{bubbles}(c) illustrates the probability of the spirals' escape as
a function of the speed of the border in a thin 3D grid of cells.
When the border moves too slow, it tends to ``drag'' the spiral waves
with it, so none penetrate into the outer zone,
as was the case in the simulation shown in panels (a,b).
When the border speed
is too high, then again no spiral waves are observed in the better
coupled upper layer, as they do not have enough time to develop. So,
as can be seen from the far right graph in \Fig{bubbles}(c), the
escapes are possible when the transitional middle layer moves faster
than a typical velocity of a spiral wave drift, but slower than the
conduction velocity (both speeds are measured for the conditions of
suppressed coupling which can be found in the border
zone).
In particular, the maximal number of spirals was observed
at the border speed of $\c=4\,\cell/\s$. 
A snapshot half way through a simulation with $\c=3\,\cell/\s$, with a few
spirals that have already penetrated the outer zone, is shown on the
leftmost panel.
As we noted earlier, in reality the border zone
 speed may vary in a very broad range.

Simple escape into the well-coupled upper layer is not enough for the
scrolls to cause fibrillation in 3D.  Scroll waves are typically born
as ``scroll rings'' with closed filaments.  As shown above, in our
model, negative filament tension is predicted by the theory for the
smaller values of excitability parameter $\param$. Moreover, the
effects of filament tension of either sign can be obstructed by the
discrete structure of the model tissue, which is particularly
essential in the conditions of the suppressed coupling.

\titfigure{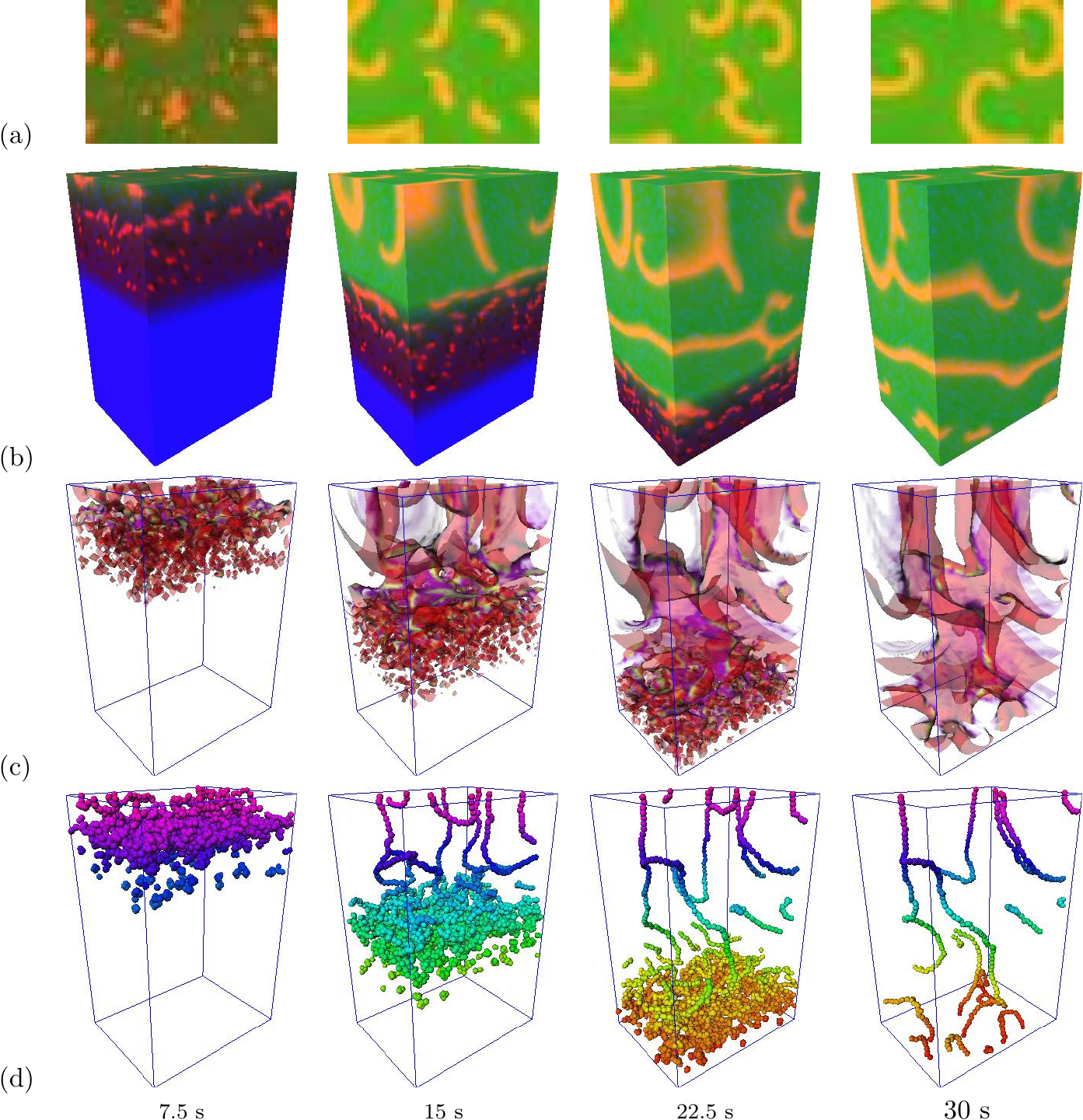}{
  Moving border zone in 3D: vortex formation.
}{
  ``Toy'' set of parameters: $\Dmax=5\cdot10^{-5}\,\cm^2/\s$,
  $\parmax=0.13$, $\delpar=0.1$, box size
  $40\times30\times60$ cells,
  border speed $\c=3\,\cell/\s$.
  Left to right: successive moments of time. 
  (a) Activation patterns at the top face of the box.
  (b) 3D view of activation patterns at the surfaces of the box.
  (c) Excitation fronts as semi-transparent surfaces. 
  (d) Vortex filaments visualized as phase singularities where 
  simultaneously $V=-35\,\mV$ and $f=0.85$.
  See also the supplementary video~\texttt{fig12.mpg}.
}{scrolls}

\Fig{scrolls} presents a simulation with the ``toy'' set of
parameters in a box size $40\times30\times60$ cells and a 
higher border speed of $\c=3\,\cell/\s$,
when moving
border zone led to generation of multiple scrolls which stayed in the
medium after the zone was gone
(see also the supplementary video~\texttt{fig12.mpg}).
\Fig{scrolls}(a) shows the top view of the
3D box at four selected instants. To a viewer this will appear as
small ectopic sources developing into larger spiral waves. 
\Fig{scrolls}(b)
reveals the underlying 3D waves as they would look on the 
side
faces of
the box. The 3D scrolls originate deep within the poorly coupled
layers of the ischaemic tissue and are spreading upwards as the border
zone moves downwards. 
\Fig{scrolls}(c)
shows in transparent colors the
wavefronts of these newly born scrolls.  Finally, 
\fig{scrolls}(d)
shows scroll filaments visualized as phase singularities defined
as the points where simultaneously $V=-35\,\mV$ and $f=0.85$. The
dense cloud of the singularities corresponds to the area where the
microscopic heterogeneities cause multiple wavebreaks.  Some of them
develop into fully fledged scroll waves, which do not collapse and
spread through the whole network of cells to instigate persistent,
self-supporting fibrillatory activity.  Note that at the value of
$\param$ used here, the filament tension is positive, and the scrolls
in the upper layer would tend to collapse were it not for the 
effect of the medium discreteness which according to~\fig{ft}(c,d) is
very essential at this
artificially low value of  $\Dmax=5\cdot10^{-5}\,\cm^2/\s$.
This simulation 
shown in~\fig{scrolls}
confirms the main
conclusion based on the two-dimensional tissue culture experiments and
simulations~\cite{Pumir-etal-2005,Biktashev-etal-2008}. That is that
the key factors of the ischaemic border zone, such as the gradient of
coupling strength together with the microscopic heterogeneity and
macroscopic gradient of excitability, generate organizing centres of
sub-millimeter scale, which then penetrate into the bulk of the well
coupled tissue, where the re-entry reaches macroscopic scales.

\titfigure{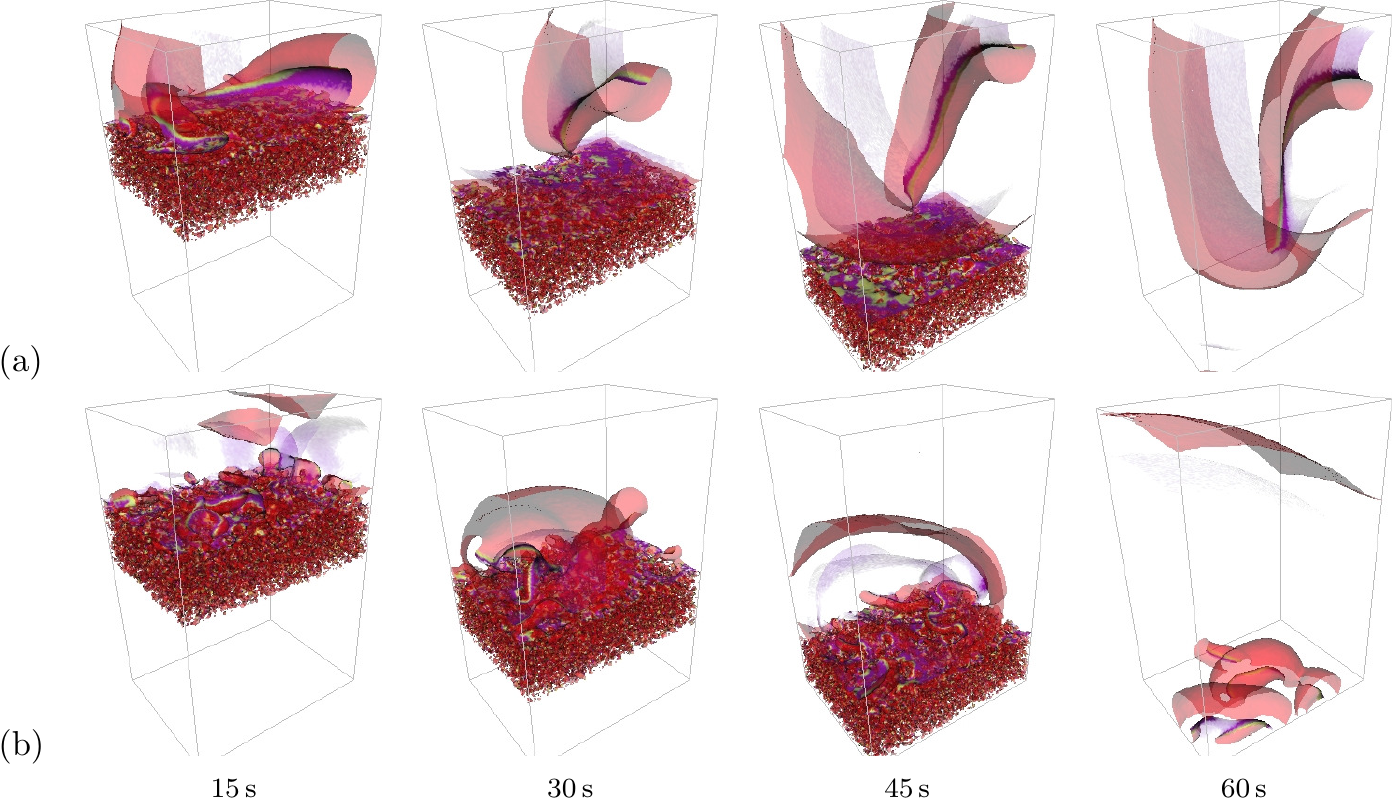}{
  Vortex formation by moving border zone.
}{
  ``More realistic'' sets of parameters: 
  $\Dmax=10^{-3}\,\cm^2/s$, $\delpar=0.5$, box size
  $120\times90\times180$ cells,
  border speed $\c=3\,\cell/\s$.
  3D views of activation patterns as in \fig{scrolls}(c): (a) $\parmax=0.105$; 
  (b) $\parmax=0.115$. 
  See also the supplementary video~\texttt{fig13.mov}.
}{large-D}

This main conclusion is supported and reinforced by simulations at
larger, more realistic values of $\Dfac$, shown in \fig{large-D}
(see also the supplementary video~\texttt{fig13.mov}).
Stronger coupling results in stronger effective averaging of the
microscopic heterogeneities. Hence, for the more realistic 3D
similutions, we have increased both the coupling strength $\Dfac$ and
the microscopic heterogeneity $\delpar$.  At this bigger $\Dmax$
value, the filament tension is already essential, as evidenced by
\fig{ft}(a).  We have chosen two values of $\param$ which correspond
to a negative and a positive tension of the generated vortex filaments
(cf~\fig{alp-graphs}(a)).  The upper row in \fig{large-D} shows
results of simulations with a negative filament tension
($\param=0.105$). In this simulation, the scroll that penetrated the
bulk of the tissue has persisted after the ischemic border zone had
disappeared. On the contrast, the lower row in \fig{large-D} shows
that for $\param=0.115$, when the scrolls in the upper layer had a
positive filament tension, they did not persist, but moved together
with the moving border zone. Continuation of the simulation
\fig{large-D}(b) led to complete elimination of all activity (not
shown).  All that is in full agreement with what could be expected
from the predictions of the asymptotic theory.

\section*{Discussion}

We have considered the quantitative predictions of the asymptotic
theory for the forces acting on rotating waves of activity that can
form within a recovering ischaemic border.  The direct numerical
simulations with deliberately arranged conditions confirmed the
theoretical predictions for the evolution of the vortices. Now, we can
answer the specific questions posed in the Introduction as follows.
\begin{enumerate}
\item \emph{``In both experiments and numerical simulations, spiral waves
were not static within the border zone. What determines the
components of the drift velocity, and why the spiral cores
can be dragged together with the moving border zone?%
''}
  \begin{itemize}
  \item The theoretical analysis of the acting forces shows that
    regions with suppressed excitability $\param$ are attracting for
    spirals, both if applied as a smooth gradient, or as a localized
    heterogeneity. Conversely, if the upper layer of the
    boundary layer has a higher excitability, it tends to repel
    spirals. This implies dragging the cores of the newly born re-entries
    by the moving transitional border zone down towards the bottom
    layer with the lowered excitability, and preventing them from
    escaping into the upper layer and ultimately into the 
    normal tissue with higher excitability.
  \item 
    At the relatively low values of excitability $\param$ in the upper layer
    corresponding to $b_2=\Re{\senscurv}<0$, the spirals are repelled
    from the transitional border layer into the better coupled upper
    layer with higher diffusivity.  This case corresponds to the
    simulation with $\param=0.105$ shown in \fig{large-D}(a). In this
    simulation, the newly born scroll penetrated the bulk of the tissue
    and persisted even after the recovering border zone ceased to
    exist.

    At the relatively high values of excitability $\param$ in the upper layer
    corresponding to $b_2=\Re{\senscurv}>0$, a gradient of diffusivity
    drives spiral waves towards areas of smaller diffusivity, \ie\ towards
    the poor coupled bottom layer. This case corresponds to the simulation
    with $\param=0.115$ shown in \fig{large-D}(b). In this simulation, the
    newly born scroll filaments never managed to get
    far into the upper layer,
    where the positive filament tension further helped to complete
    their elimination.   
   
    Hence, a relatively high excitability in the
    upper layer will suppress the transition to 
    fibrillatory-like state for
    \emph{two} reasons: the gradient of excitability will prevent 
    the cores of spirals or filaments of scrolls from escaping into
    the more excitable outer zone; and at higher excitability, the
    gradient of coupling will also drag them away from the better
    coupled outer zone.
    
  \end{itemize}
\item \emph{``In both experiments and numerical simulations, the drift of the
spirals was interrupted by their ``pinning'' to clusters of cells. We
have shown numerically that these can be cell clusters of either
elevated or suppressed excitability. What is the mechanism of such
pinning?%
''}
  \begin{itemize}
  \item The theoretical analysis shows that a combination of
    acting forces generated by smooth gradients of tissue properties and a
    localized inhomogeneity in excitability parameter $\param$ may lead to
    temporary or permanent pinning of drifting spirals. The chances of
    pinning depend on the trajectory of the drifting spiral and geometry
    of the heterogeneity, and it may happen at either sign of the
    inhomogeneity (\ie\ locally increased or decreased excitability).
  \item There is more than one mechanism of pinning. Apart from
    pinning to an attracting inhomogeneity, the drift can also be stopped
    by a certain spatial arrangement of repelling inhomogeneities. Even if
    ``permanent'' pinning is not achieved, a temporary pinning still may be
    observed for some finite time if the trajectory of the spiral core
    passes near an unstable equilibrium. There is also a theoretical
    possibility of ``orbital motion'' which however is not realized in the
    present model at interesting values of parameters.
  \end{itemize}
\item \emph{``In both experiments and numerical simulations, the episodes of
spiral drift and pinning alternated. What is the mechanism by which
pinning can give way to further drift?%
''}
  \begin{itemize}
  \item Correspondingly, there is more than one mechanism of
    unpinning.  One is that due to the border zone dynamics, parameters of
    the tissue may change in such a way that gradient-induced force
    exceeds the tear-off threshold. The other is that the spiral wave
    core drifts away from the pinning site because its position
    there was unstable in the first place.
  \end{itemize}
\item \emph{``One of arrhythmogenic scenarios proposed
  in~\cite{Pumir-etal-2005,Biktashev-etal-2008} involved pinning of a
spiral wave to a local heterogeneity which persists long enough until
the border zone passes and the spiral gets into the better coupled
tissue. Is this scenario viable in 3D?%
''}
  \begin{itemize}
  \item In 3D, in addition to whatever dynamics 2D spiral waves might
    have, scroll waves exhibit additional dynamics associated with the
    motion of filament, and characterized by the \emph{filament
      tension} and the {\em binormal drift coefficient}.
    In the considered tissue model, the filament tension is small
    compared to the binormal drift coefficient, and changes sign in
    the relevant range of excitability parameter. This means that
    scroll waves that managed to escape into the well coupled upper
    zone, might not necessarily immediately collapse.
  \item The scroll filaments that managed to stay until the tissue is
    recovered, may not collapse but survive, if the filament tension
    is negative. These filaments may subsequently generate scroll wave
    turbulence. Note a nontrivial coincidence following from the
    asymptotic theory: excitability of the upper layer at the lower
    range of parameter $\param$ ensures the negative filament tension
    and hence is a condition of survival of scrolls in that zone, and
    it also ensures that the specific force caused by the coupling
    gradient repells the scrolls into the upper, better coupled layer.
    So here we have a 
    \emph{third} reason a relatively high
    excitability in
    the outer layer is ``anti-arrhythmic'': at higher excitability, the
    scrolls in the outer layer are less likely to survive due to
    3D effects.
  \item Further, there are some features revealed by the 2D
    simulations which are beyond direct applicability of the
    asymptotic theory. That is the effect of the dicreteness of the
    medium, which particularly matters at low values of
    diffusivity. The discretness of the medium can arrest the drift of
    spiral cores, and when applied to 3D scrolls, the filaments
    can freeze as long as their curvature is not too high, and the
    ``filament tension'' component of their drift freezes sooner than
    the ``lateral binormal drift'' component.  Therefore, the scroll
    filaments that managed to stay until the tissue is recovered, may
    not collapse but survive, as their filament tension is frozen due
    to low diffusivity. In that case of the ``frozen'', zero filament
    tension, the regime might rather look like a persistent
    tachycardia similar to the pinned 2D spiral regime.
  \end{itemize}
\end{enumerate}

To summarize, we explored a biophysically plausible mechanism as to
how ectopic beats and spreading scrolls of abnormal activity can be
generated from the recovering boundary of acutely ischaemic tissue.
Complex boundary behaviour in heterogeneous cell network was modeled
with certain assumptions and simplifications, extensively discussed in
our previous publications~\cite{Pumir-etal-2005,Biktashev-etal-2008}.

With all the assumptions and limitations, the following combined
conclusions can be made based on the \invitro\ and \insilico\ data
from our previous publications and the current study.  First, the data
suggested that the combination of the two gradients (\ie\ the spatial
gradient in cell-to-cell coupling and the temporal gradient in
excitability/automaticity) ensured that somewhere within the border
zone there was a region where multiple ectopic sources were
continuously being formed. They were highly localized focal points of
activity, with activation spreading only to a few surrounding
cells. Number of ectopic sources and specific window of conditions
when they occured were affected by the degree of the network
heterogeneity. Secondly, the data argued that if the ectopically
active layer was sufficiently wide and/or the overall cell
automaticity rose, ectopic sources developed into target-like
waves. If coupling gradient and automaticity levels remained
spatiotemporally fixed, the pattern of target-like sources persisted
and no spiral activity was observed. However, when cell automaticity
rose and/or border zone moved in space, the propagation patterns
became non-stationary. This led to multiple wavebreaks resulting in
spiral generation activity. The spiral waves typically demonstrated
start-stop drifting behaviour, as a result of competing forces between
pinning force due to local heterogeneity and a gradient-induced
directional drift. The likelihood of a spiral escape into the better
coupled upper tissue zone depend on the speed at which the border zone
moves in space.

Our extrapolation of 2D events into 3D is more
  theoretical, as tissue culture experiments similar to
  those described in~\cite{Arutunyan-etal-2003,Pumir-etal-2005} are
  not feasible in 3D. Still, this extrapolation
has shown that the border zone can
give rise to 3D analogues of spirals, the scroll waves. If a scroll
wave escapes into a better coupled tissue it will not necessarily
cause fibrillation, because the scroll wave with positive filament
tension have tendency to collapse. However, our simulations have shown
that this collapse of newly generated scrolls is not inevitable and,
instead, scroll filaments can stabilise or, in case of negative
filament tension, expand and multiply leading to a fibrillation-like
state.

In this study, we considered the asymptotic theory's quantitative
predictions for the forces acting on a cardiac re-entry, and causing
its drift, in the vicinity of the ischaemic border zone. The
theoretical predictions allow to tell apart and highlight different
mechanisms of arrythmogenesis by the ischaemic boder zone in
three-dimentiontional settings.  The direct numerical simulations with
deliberately arranged conditions confirmed the theoretical predictions
for the drift.

We fully realize that in vivo, the above considered scenarios will be
affected by multiple additional factors. These might include 
excitability kinetics different from the simplified
  generic model we used here,
presence
of highly excitable Purkinje fibers, macroscopic myofiber orientation,
coronary vessels, fibrous or fat deposits, transmural differences in
myocytes metabolic activity and their sensitivity to ischaemia. Yet,
with all its limitations, this study represents one of the first
attempts to theoretically explore a very complex set of highly
arrhythmogenic conditions that can occur on the boundary of the
recovering ischaemic tissue.

\section*{Acknowledgments}
We thank A. Arutunyan for experimental and conceptual input that led
  to this work. The experimental studies that served as a foundation
  for this work were supported by the National Institutes of Health
  (HL076722, HL095828). The response functions calculation technology and
  DXSpiral software was co-authored by D.~Barkley, G.~Bordyugov and
  A.~Foulkes and partly supported by Engineering and Physical Sciences
  Research Council, UK (EP/D074789/1). The QUI/Beatbox numerical
  simulation software was co-authored by A.~Karpov and R.~McFarlane
  and partly supported by Engineering and Physical Sciences Research
  Council, UK (EP/I029664/1).



\newpage

\section*{Supporting information}

\begin{itemize}
\item File \texttt{appendix.pdf}: Appendix, describing details of the
  Beeler-Reuter-Pumir kinetic model, and of our numerical scheme. 
\item File \texttt{fig12.mpg}: Movie illustration to figure 12. 
\item File \texttt{fig13.mpg}: Movie illustration to figure 13. 
\end{itemize}

\end{document}